\documentclass[%
 reprint,
superscriptaddress,
 amsmath,amssymb,
 aps,
]{revtex4-2}

\usepackage{graphicx}
\usepackage{dcolumn}
\usepackage{bm}
\usepackage{color}
\usepackage[colorlinks,linkcolor=blue,citecolor=blue]{hyperref}
\usepackage{bm}
\usepackage{amsmath}
\usepackage{float}

\begin{document}

\preprint{APS/123-QED}

\title{Enhanced coherent light-matter interaction and room-temperature quantum yield of plasmonic resonances engineered by a chiral exceptional point}

\author{Yuwei Lu}
\email[Corresponding Author: ]{luyw5@fosu.edu.cn}
\affiliation{School of Physics and Optoelectronics, South China University of Technology, Guangzhou 510641, China}
\affiliation{School of Physics and Optoelectronic Engineering, Foshan University, Foshan 528000, China}

\author{Haoxiang Jiang}%
\affiliation{Guangdong Provincial Key Laboratory of Quantum Engineering and Quantum Materials, School of Physics, South China Normal University, Guangzhou 510006, China}%
\affiliation{Guangdong-Hong Kong Joint Laboratory of Quantum Matter, Frontier Research Institute for Physics, South China Normal University, Guangzhou 510006, China}%

\author{Renming Liu}
\email[Corresponding Author: ]{liurm@henu.edu.cn}
\affiliation{School of Physics and Electronics, International Joint Research Laboratory of New Energy Materials and Devices of Henan Province, Henan University, Kaifeng 475004, China}%
\affiliation{Institute of Quantum Materials and Physics, Henan Academy of Sciences, Zhengzhou 450046, China}%



\begin{abstract}
Strong dissipation of plasmonic resonances is detrimental to quantum manipulation. To enhance the quantum coherence, we propose to tailor the local density of states (LDOS) of plasmonic resonances by integrating with a photonic cavity operating at a chiral exceptional point (CEP), where the phase of light field can offer a new degree of freedom to flexibly manipulate the quantum states. A quantized few-mode theory is employed to reveal that the LDOS of the proposed hybrid cavity can evolve into sub-Lorentzian lineshape, with order-of-magnitude linewidth narrowing and additionally a maximum of eightfold enhancement compared to the usual plasmonic-photonic cavity without CEP. This results in the enhanced coherent light-matter interaction accompanied by the reduced dissipation of polaritonic states. Furthermore, a scattering theory based on eigenmode decomposition is present to elucidate two mechanisms responsible for the significant improvement of quantum yield at CEP, the reduction of plasmonic absorption by the Fano interference and the enhancement of cavity radiation through the superscattering. Importantly, we find the latter allows achieving a near-unity quantum yield at room temperature; in return, high quantum yield is beneficial to experimentally verify the enhanced LDOS at CEP by measuring the fluorescence lifetime of a quantum emitter. Therefore, our work demonstrates that the plasmonic resonances in CEP-engineered environment can serve as a promising platform for exploring the quantum states control by virtue of the non-Hermiticity of open optical resonators and building the high-performance quantum devices for sensing, spectroscopy, quantum information processing and quantum computing.
\end{abstract}

\maketitle


\section{Introduction}\label{sec1}

Plasmonic nanocavities have achieved great success in realizing the room-temperature strong coupling between the localized surface plasmon resonance and a few or even single quantum emitter (QE) at the nanoscale \cite{RN1,RN2,RN3,RN4}. This leads to the formation of hybrid plasmon-QE states, usually termed plexcitons, that enables a rich variety of advanced quantum technologies, such as ultrafast single-photon source \cite{RN5,RN6,RN7}, single-qubit control \cite{RN8,RN9,RN10}, and ultrasensitive quantum sensing \cite{RN11,RN12}. Nevertheless, to move toward practical applications, plexcitonic systems still call for further engineer to prolong the coherence time, which is severely limited by the large Ohmic loss in metals. To overcome this challenge, the oscillating photonic building blocks like photonic crystal and whispering-gallery-mode (WGM) cavities, have been employed to reduce the plasmonic dissipation while simultaneously produce larger cooperativity of light-matter interaction \cite{RN13,RN14,RN15,RN16,RN17,RN18,RN19,RN20,RN21,RN22}. This attributes to the constructive interference of scattering paths in hybrid cavity with red-detuned plasmon-photon coupling, which provides the unique benefit of higher Purcell factor than separate entities \cite{RN14,RN15,RN18,RN23}. Such plasmonic-photonic hybrids also emerge as one of successful approaches to manipulate quantum states. In this context, on resonance the plasmonic absorption can be suppressed by exploiting the principle of bound states in the continuum \cite{RN24} and as a result, bound polaritonic states with anomalously small decay can form \cite{RN25}. Therefore, previous studies have shown the ability of photonic elements to tailor the electromagnetic environment of LSPR and hence modify the plasmon-QE interaction. 

Leakage of optical resonators, often considered detrimental to quantum devices, is inevitable but at the same time offers new avenues for light manipulation exploiting the non-Hermitian degeneracies, known as exceptional points (EPs) \cite{RN26,RN27,RN28}. With the flourish of non-Hermitian quantum mechanics, the dimensionality reduction of state space at EPs arises as a new strategy to steer the quantum states and spurs the exploration of quantum-optics applications, including tunable photon statistics \cite{RN29,RN30}, enhanced sensitivity of quantum sensing \cite{RN31,RN32,RN33}, and emission control of quantum light \cite{RN34,RN35,RN36}, to name a few. Nevertheless, the isolated EPs are very sensitive to the external perturbations, this feature hinders the access to EPs in quantum systems. For this reason, extending isolated EPs to higher dimension has emerged as an alternative for the observation of related phenomena \cite{RN37,RN38,RN39,RN40}. Particularly, a chiral EP (CEP), which is a two-dimensional collection of second-order EPs in parameter space, can be implemented in WGM cavity by introducing the unidirectional coupling between the degenerate counterclockwise (CCW) and clockwise (CW) modes \cite{RN39,RN41,RN42,RNPRL2014}. This structure exhibits a fascinating feature that the CEP hosting in WGM cavity is immune to the inevitable experimental uncertainties and fabrication imperfections, thus can offer a robust platform for exploring the non-Hermitian physics. 

Inspired by these pioneering works, we propose to bring the light-matter interaction into the strong coupling regime by engineering the local density of states (LDOS) of a plasmonic nanocavity through the integration with a WGM photonic cavity featuring CEP, which we call hybrid CEP cavity hereafter, while the usual plasmonic-photonic cavity without CEP is called hybrid cavity. A full quantum model is first built to reveal that the LDOS of plasmon-photon hybrid mode is related to the phase of light field at CEP, and can be tailored to further enhance while at the same time the linewidth is significantly reduced. It also reveals the mechanisms of order-of-magnitude enhancement of quantum yield at CEP, by either reducing the plasmonic absorption or enhancing the photonic radiation. Combined with the electromagnetic simulations of a simple design of hybrid CEP cavity, we predict that high quantum yield at room temperature can be achieved in realistic structures with proper design. 

The paper is organized as follows. In Sec. \ref{sec2}, we propose a prototypical mode of hybrid plasmonic-photonic cavity with a CEP hosting in WGM cavity, and formulate its LDOS based on the few-mode quantization description, which allows to analytically express LDOS through the frequencies, decay rates and coupling rates of uncoupled plasmonic and photonic modes. Sec. \ref{sec3} investigates the modification of LDOS at CEP, where the phase of light field offers a new degree of freedom to flexibly tune the magnitude and linewidth of LDOS. Subsequently, in Secs. \ref{sec4} and \ref{sec5} we demonstrate the advantages of hybrid CEP cavity in enhancing the coherent light-matter interaction and quantum yield, respectively, as two examples of quantum-optics applications. The formalism of LDOS is applied to a realistic structure in Sec. \ref{sec6}, and analyze the enhanced light-matter interaction and quantum yield by varying the parameters of structure geometry, from which we summarize the design rules for hybrid CEP cavity to achieve high quantum yield at room temperature. Finally, we conclude in Sec. \ref{sec7}. 

\begin{figure}[t]
\centering\includegraphics[width=0.88\linewidth]{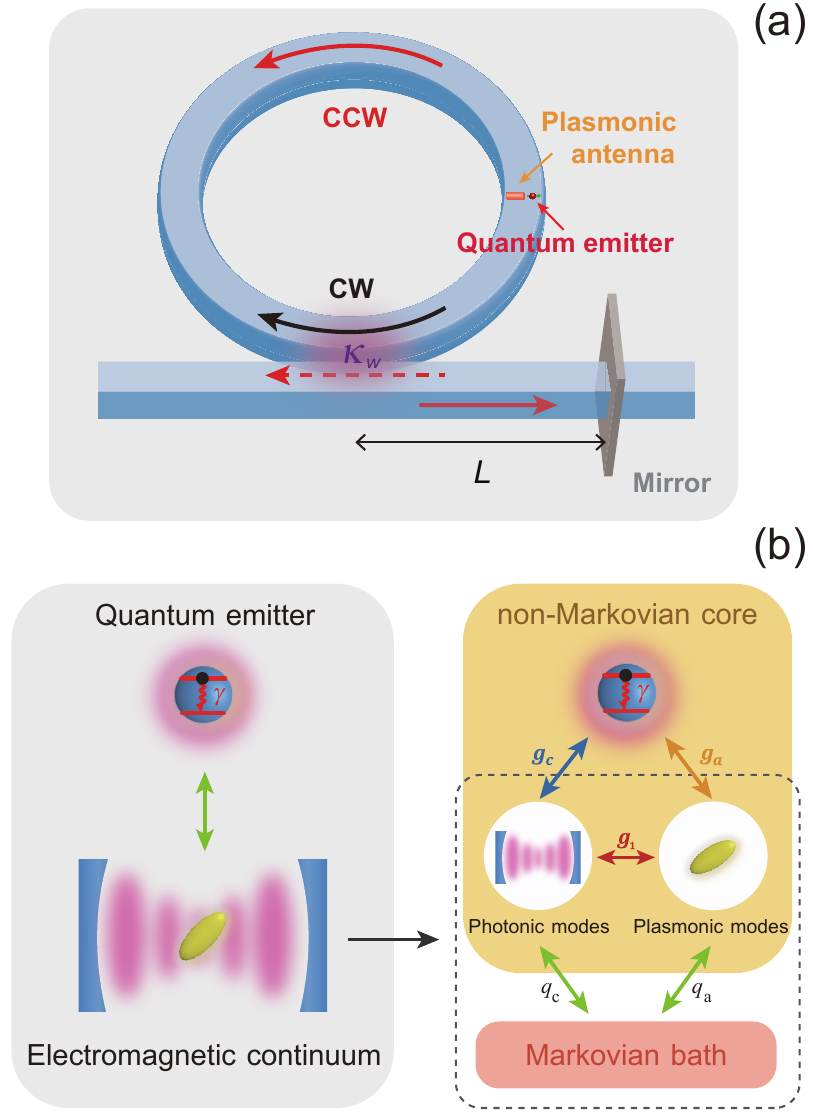}
\caption{(a) Schematic diagram of a plasmonic-photonic cavity operating at a chiral exceptional point (CEP) based on whispering-gallery-mode (WGM) cavity, which supports the degenerate clockwise (CW) and counterclockwise (CCW) modes with unidirectional coupling provided by the mirror. (b) Conceptual sketch of the mapping of the local density of states of plasmonic-photonic cavity constituted of continuous electromagnetic modes into a quantized few-mode model, where the composed system can be decomposed into a non-Markovian core and a Markovian bath. The gray dashed line indicates the structured environment of plasmonic-photonic cavity. }
\label{fig1}
\end{figure}

\section{Model and theory}\label{sec2}

The hybrid CEP cavity under investigation is schematically sketched in Fig. \ref{fig1}(a), which consists of a usual plasmonic-photonic cavity and a waveguide evanescently coupled to a pair of degenerate WGM modes with opposite propagating direction, i.e., the CW and CCW modes. A CEP is constructed through the unidirectional coupling from CCW mode to CW mode generated by the mirror at the right end of waveguide. The mirror is assumed to have unity reflectivity. An effective phase $\phi_s=\beta L$ is induced for the light field of CCW mode propagating from the waveguide-cavity junction to the mirror, with $\beta$ and $L$ being the propagation constant of the waveguide and the distance between the cavity-waveguide junction and the mirror, respectively. When interacting with a QE, hybrid CEP cavity in general manifests non-Lorentzian LDOS and can be described as a structured electromagnetic environment that constitutes continuous bosonic modes in quantized representation \cite{RN44}. On the other hand, the non-Lorentzian LDOS stems from the coupling of plasmonic and photonic resonances, thus can in principle be generated using the parameters of individual resonances and the coupling rate between them. With this prospect, we develop a quantized LDOS theory for hybrid CEP cavity in this section. 

We consider a dipolar plasmonic antenna where the nonradiative higher-order modes are well separated from the dipolar mode, so that the plasmonic antenna can be treated as a single-mode resonator with resonance frequency $\omega_a$. This approximation is valid for many plasmonic antennas, such as the elongated rods \cite{RN4} and dimer structures \cite{RN23,RN45}. The dipolar plasmonic mode is coupled to both CCW and CW modes with resonance frequency $\omega_c$. Accordingly, we can obtain a quantized few-mode model of hybrid CEP cavity where the continuous bosonic environment is decomposed into a non-Markovian core and a Markovian bath \cite{RN25,RN46,RN47,RN48,RN49}, as illustrated in Fig. \ref{fig1}(b), which yields the same reduced quantum dynamics as the original system. Based on this equivalent model, an extended cascaded quantum master equation of a two-level QE interacting with hybrid CEP cavity is derived in Appendix \ref{aa}, and employed to describe the quantum dynamics of the composed cavity quantum electrodynamics (QED) system 
\begin{equation}\label{eq1}
\dot{\rho}=-i[H, \rho]+\mathcal{D}[\rho]
\end{equation}
with Lindblad operator
\begin{equation}\label{eq2}
\begin{aligned}
\mathcal{D}[\rho]= & \gamma \mathcal{L}\left[\sigma_{-}\right] \rho+\kappa_a \mathcal{L}[a] \rho+\kappa\left(\mathcal{L}\left[c_{c c w}\right] \rho+\mathcal{L}\left[c_{c w}\right] \rho\right) \\
& +\kappa_c\left(e^{i \phi}\left[c_{c c w} \rho, c_{c w}^{\dagger}\right]+e^{-i \phi}\left[c_{c w}, \rho c_{c c w}^{\dagger}\right]\right)
\end{aligned}
\end{equation}
where the total Hamiltonian reads $H=H_0+H_I$, with the free Hamiltonian $H_0=\omega_a a^{\dagger} a+\omega_c c_{ccw}^{\dagger} c_{ccw}+\omega_c c_{cw}^{\dagger} c_{cw}+\omega_0 \sigma_{+} \sigma_{-}$ and the interaction Hamiltonian $H_I=g_1 (c_{ccw}^{\dagger} a+a^{\dagger} c_{ccw} )+g_a (a^{\dagger} \sigma_{-}+\sigma_{+} a)+g_c (c_{ccw}^{\dagger} \sigma_{-}+\sigma_{+} c_{ccw} )+g_c (c_{cw}^{\dagger} \sigma_{-}+\sigma_{+} c_{cw} )$. $a$ and $c_{ccw}$ (or $c_{cw}$) are the bosonic annihilation operators for dipolar plasmonic mode and CCW (or CW) mode, respectively, while $\sigma_{-}$ is the lowering operator of QE with transition frequency $\omega_0$. $g_1$, $g_a$ and $g_c$ denote the coupling rates for plasmon-photon, plasmon-QE and photon-QE interactions, respectively. The coupling rates are assumed to be real since the mode volume of cavity is much larger than that of plasmonic antenna and hence the latter can be considered as a dipole scatter \cite{RN14,RN15}. $\mathcal{L}[O] \rho=2 O \rho O^{\dagger}-O^{\dagger} O \rho-\rho O^{\dagger} O$ is the Liouvillian superoperator for the dissipation of operator $O$. The first line of Eq. (\ref{eq2}) introduces the dissipation for individuals, where the QE decay rate is $\gamma=\gamma_0+\gamma_{nr}$, with $\gamma_0$ being the free-space emission rate of QE and $\gamma_{nr}$ accounting for the nonradiative decay rate of QE, which is typically $10-20 \text{meV}$ for QEs at room temperature \cite{RN64,RN65}. $\kappa_a=\kappa_o+\kappa_r$ stands for the total decay rate of dipolar plasmonic mode, with $\kappa_o$ and $\kappa_r$ being the radiative and nonradiative decay rates, respectively. $\kappa=\kappa_c+\kappa_i$ denotes the total decay rate of WGM modes, where $\kappa_c$ stems from the evanescent coupling of cavity to the guided mode of waveguide, which can be tuned by adjusting the cavity-waveguide separation. $\kappa_i$ is the intrinsic decay of CCW and CW modes resulting from the material absorption and the coupling of cavity to the modes other than the guided modes of waveguide. We first omit the intrinsic decay $\kappa_i$ in consideration of the high-$Q$ feature of WGM modes, but retrieve it in Sec. VI when analyzing the quantum yield of realistic structures. The second line of Eq. (\ref{eq2}) describes the cascaded interaction that the CW mode is driven by the output field from the CCW mode through the mirror reflection, where $\phi=2\phi_s$ is the roundtrip phase factor. 

For the cases of spontaneous emission and weak drive studied in this work, we restrict the states space of quantum system in single-excitation manifold, where the equations of motion can be derived from Eqs. (\ref{eq1}) and (\ref{eq2}) 
\begin{equation}\label{eq3}
\frac{d}{d t} \vec{p}_0=-i \mathbf{M}_0 \vec{p}_0
\end{equation}
with $\vec{p}_0=\left[\left\langle\sigma_{-}\right\rangle,\left\langle a \right\rangle,\left\langle c_{c c w}\right\rangle,\left\langle c_{c w}\right\rangle\right]^T$ and the matrix $\mathbf{M}_0$ being
\begin{equation}\label{eq4}
\mathbf{M}_0=\left[\begin{array}{cccc}
\omega_0 - i\frac{\gamma}{2} & g_a & g_c & g_c \\
g_a & \omega_a-i \frac{\kappa_a}{2} & g_1 & g_1 \\
g_c & g_1 & \omega_c-i \frac{\kappa}{2} & 0 \\
g_c & g_1 & -i \kappa_c e^{i \phi} & \omega_c-i \frac{\kappa}{2}
\end{array}\right]
\end{equation}
We then change the basis of WGM cavity into the representation of standing wave modes \cite{RN51}
\begin{equation}\label{eq5}
c_{1}=\frac{1}{\sqrt{2}}\left(c_{ccw}+c_{cw}\right), \quad c_{2}=\frac{1}{\sqrt{2}}\left(c_{ccw}-c_{cw}\right)
\end{equation}
Substituting Eq. (\ref{eq5}) into Eq. (\ref{eq3}), we obtain $d\vec{p}/dt=-i \mathbf{M}_p \vec{p}$, with $\vec{p}=\left[\left\langle\sigma_{-}\right\rangle,\left\langle\ a \right\rangle,\left\langle c_1\right\rangle,\left\langle c_2\right\rangle\right]^T$. The matrix $\mathbf{M}_p$ is given by
\begin{equation}\label{eq6}
\mathbf{M}_p=\left[\begin{array}{cccc}
\omega_0 - i\frac{\gamma}{2} & g_a & \sqrt{2}g_c & 0 \\
g_a & \omega_a-i \frac{\kappa_a}{2} & \sqrt{2}g_1 & 0 \\
\sqrt{2}g_c & \sqrt{2}g_1 & \omega_c-i \frac{\kappa_{+}}{2} & i \frac{\kappa_c}{2} e^{i \phi} \\
0 & 0 & -i \frac{\kappa_c}{2} e^{i \phi} & \omega_c-i \frac{\kappa_{-}}{2}
\end{array}\right]
\end{equation}
where $\kappa_{\pm}=\kappa_i + \kappa_c(1 \pm e^{i \phi})$. We can see that without the mirror (i.e., $\kappa_c=0$), the standing wave mode $c_2$ becomes uncoupled and Eq. (\ref{eq6}) returns to the single-mode treatment of WGM cavity alternatively used in the literature \cite{RN51,RN52,RN53}. In the presence of mirror, $c_2$ is still decoupled from the QE and plasmon, which can simplify the subsequent derivation of LDOS.

\section{Local density of states of hybrid CEP cavity}\label{sec3}

The normalized LDOS, i.e., Purcell factor, is linked to the spectral density $J(\omega)$ through the relation $P(\omega)=J(\omega)/J_0 (\omega)$, with $J_0 (\omega)=\omega^3 \mu^2/6\pi^2 \hbar\epsilon_0 c^3$ being the spectral density of a QE with dipole moment $\mu$ \cite{RN50} in the free space, where $c$ is the speed of light and $\epsilon_0$ is the permittivity of vacuum. The spectral density of hybrid CEP cavity is given by \cite{RN46,RN47}
\begin{equation}\label{eq7}
J(\omega)=\operatorname{Re}\left[i \chi_{sys}\left(\omega\right)\right]=\operatorname{Re}\int_{-\infty}^{+\infty} d \tau e^{i \omega \tau} \left\langle \Lambda^{\dagger}(0) \Lambda(\tau)\right\rangle
\end{equation}
with 
\begin{equation}\label{eq8}
\Lambda(t) = g_a a(t) + \sqrt{2}g_c c_{1}(t)
\end{equation}
where $\chi_{sys}\left(\omega\right)$ defines the polarizability of hybrid CEP cavity. We can see that the spectral density can be separated into three parts, i.e., $J(\omega)=J_a (\omega)+J_c (\omega)+J_{ac} (\omega)$, with $J_a(\omega)=g_a^2 \operatorname{Re}\left\{\mathcal{F}\left[\left\langle a^{\dagger}(0) a(\tau)\right\rangle\right]\right\}$ and $J_c(\omega)=2g_c^2 \operatorname{Re}\left\{\mathcal{F}\left[\left\langle c_1^{\dagger}(0) c_1(\tau)\right\rangle\right]\right\}$ being the modified plasmon and cavity response, respectively. $J_{ac}(\omega)=\sqrt{2}g_a g_c \operatorname{Re}\left\{\mathcal{F}\left[\left\langle a^{\dagger}(0) c_1(\tau)\right\rangle\right]\right\}$ contains the interference between the plasmon and cavity, where $\mathcal{F}\left[ \cdot \right]$ represents the Fourier transform. The two-time correlation functions $\left\langle a^{\dagger}(0) a(\tau)\right\rangle$, $\left\langle c_1^{\dagger}(0) c_1(\tau)\right\rangle$ and $\left\langle a^{\dagger}(0) c_1(\tau)\right\rangle$ can be calculated using the quantum regression theorem \cite{RN50}. Taking $\left\langle a^{\dagger}(0) a(\tau) \right\rangle$ as an example, its dynamics follows the equation
\begin{equation}\label{eq9}
\frac{d}{d \tau} \left\langle a^{\dagger}(0)\vec{c}(\tau) \right\rangle=-i \mathbf{M}_s \left\langle a^{\dagger}(0)\vec{c}(\tau) \right\rangle
\end{equation}
where $\vec{c}(\tau)=\left[\left\langle a(\tau) \right\rangle,\left\langle c_1(\tau) \right\rangle,\left\langle c_2(\tau) \right\rangle\right]^T$. The matrix $\mathbf{M}_s$ takes the form 
\begin{equation}\label{eq10}
\mathbf{M}_s=\left[\begin{array}{ccc}
\omega_a - i \frac{\kappa_a}{2} & \sqrt{2}g_1 & 0 \\
\sqrt{2}g_1 & \omega_c-i \frac{\kappa_{+}}{2} & i \frac{\kappa_c}{2} e^{i \phi} \\
0 & -i \frac{\kappa_c}{2} e^{i \phi} & \omega_c-i \frac{\kappa_{-}}{2}
\end{array}\right]
\end{equation}
With initial condition $\left\langle a^{\dagger}(0) a(0) \right\rangle = 1$, $\left\langle a^{\dagger}(0) c_1(0)\right\rangle = 0$ and $\left\langle a^{\dagger}(0) c_2(0)\right\rangle = 0$, the equation can be solved through the Fourier transform. Other correlation functions can be obtained with the similar fashion (see Appendix \ref{ab} for details). Note that for hybrid cavity, $g_a$ is in general larger than $g_c$ by over two orders of magnitude since the QE is often placed outside instead of embedded in the WGM cavity \cite{RN14,RN16,RN23,RN55}; therefore, the contributions of $J_c (\omega)$ and $J_{ac} (\omega)$ to the spectral density can be omitted, i.e., $J(\omega) \approx J_a (\omega)$. Furthermore, the decoupling of QE from WGM cavity can also facilitate to clarify the role of CEP in engineering the plasmonic resonance. Based on these reasons, in the following discussion  we take $g_c=0$ unless special noted. In this circumstance, the analytical expression of spectral density is given by 
\begin{equation}\label{eq11}
J(\omega)=-g_a^2\operatorname{Im}\left[\frac{\chi_a(\omega)}{1-g_1^2 \chi_a(\omega) \chi_{E P}(\omega) }\right]
\end{equation}
where the polarizabilities of uncoupled plasmonic antenna and CEP cavity are characterized by $\chi_a (\omega)=\left[\omega-\omega_a+i\kappa_a/2\right]^{-1}$ and $\chi_{EP} (\omega)=2\chi_c (\omega)-i\kappa_c e^{i \phi} \chi_c^2 (\omega)$, respectively, with $\chi_c (\omega)=\left[\omega-\omega_c+i\kappa_c/2\right]^{-1}$ being the polarizability of bare WGM modes. Both $-\operatorname{Im}\left[\chi_a (\omega)\right]$ and $-\operatorname{Im}\left[\chi_c (\omega)\right]$ present the conventional Lorentzian lineshape, while the squared Lorentzian term $\propto\chi_c^2 (\omega)$ in $\chi_{EP} (\omega)$ is a hallmark of second-order EP, where the relative phase $\phi$ provides an extra degree of freedom to tailor LDOS. 
\begin{figure}[t]
\centering\includegraphics[width=0.98\linewidth]{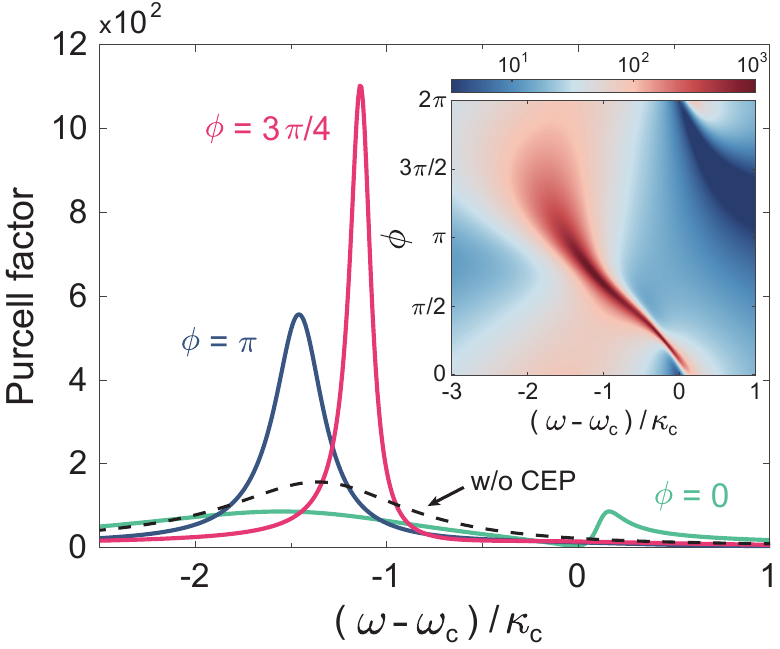}
\caption{Purcell factor of hybrid CEP cavity for various $\phi$ (solid lines). The inset shows the logarithmic plot of the Purcell factor versus frequency and $\phi$. Purcell factor of the hybrid cavity without CEP is shown for comparison (dashed black line). The parameters are $\kappa_i= \gamma_{nr}=0$, $g_1=-20\mathrm{meV}$, $g_a=10\mathrm{meV}$, and the plasmon-photon detuning $\Delta_{ac}=\omega_a-\omega_c=1\mathrm{eV}$. The quality factors of dipolar plasmonic antenna and WGM modes are $Q_a=18$ and $Q_c=2 \times 10^3$, respectively.  }
\label{fig2}
\end{figure}

\begin{figure*}[t]
\centering\includegraphics[width=0.98\linewidth]{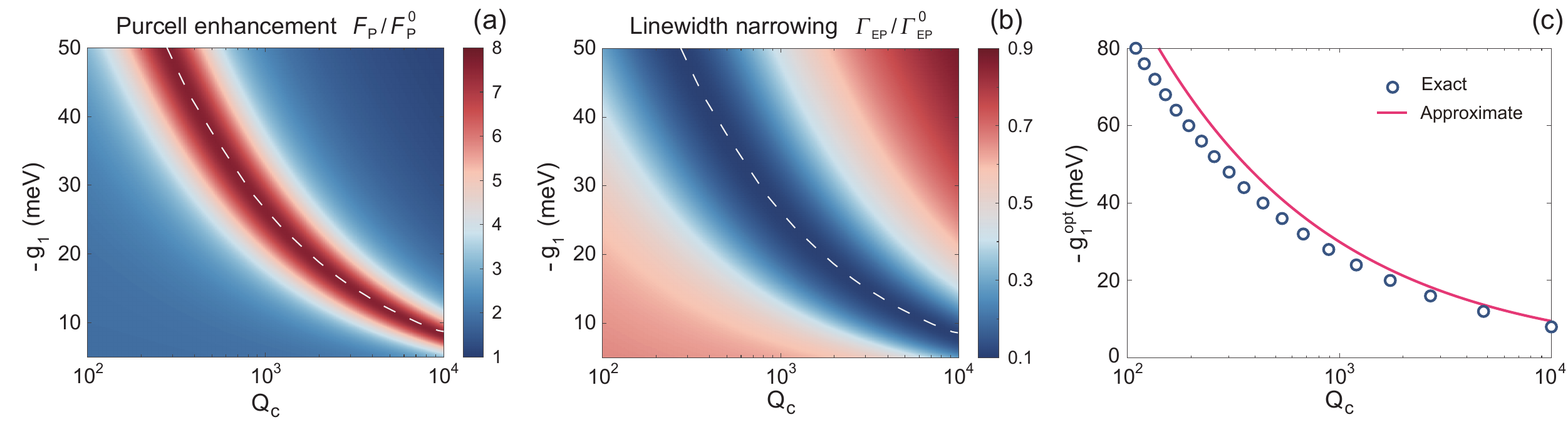}
\caption{Purcell enhancement $F_p/F_p^0$ (a) and linewidth narrowing $\Gamma_{EP}/\Gamma_{EP}^0$ (b) of hybrid CEP cavity versus $Q_c$ and $g_1$. The white dashed line traces the locations of maximal Purcell enhancement. (c) Comparison of the optimal plasmon-photon coupling $g_1^{\text{opt}}$ for the maximal Purcell enhancement obtained from the numerical results (circles) and predicted by the analytical expression $g_1^{\text{opt}}=-\sqrt{3\Delta_{ac} \kappa_c }/2$ (solid line), where $\Delta_{ac}=\omega_a-\omega_c$ is the plasmon-photon detuning. Parameters not mentioned are the same as Fig. \ref{fig2}.}
\label{fig3}
\end{figure*}
We first consider the case of red-detuned plasmon-photon interaction, with coupling rates $g_1=-20\text{meV}$ and $g_a=10\text{meV}$. This value of $g_1$ is similar to the theoretical values previously reported in the literature \cite{RN25,RN48}, while $g_a=10\text{meV}$ is far below the achievable plasmon-QE coupling rate in experiments, which can exceed 40meV for molecule QEs \cite{RN1,RN4,RN16} and is over $100\text{meV}$ for semiconductor quantum dots \cite{RN3,RN56}. Therefore, the aforementioned parameters are attainable in realistic structures. In the inset of Fig. \ref{fig2}, we plot the Purcell factor of hybrid CEP cavity versus frequency, where it demonstrates distinct lineshapes as $\phi$ varies. Fig. \ref{fig2} shows this dependence of Purcell factor on $\phi$ more explicitly. We can see that the frequency, magnitude, and linewidth of Purcell factor varies greatly with $\phi$. For example, two peaks appearing in the Purcell factor for $\phi\sim0$, with a dip around cavity resonance resulting from the deconstructive interference between the CCW and CW modes at CEP. On the contrary, Purcell factor with $\phi=\pi$ presents a single peak close to the resonance frequency of hybrid cavity (i.e., without CEP), but with a linewidth narrower than the corresponding Lorentzian function with the same maximum, demonstrating a sub-Lorentzian lineshape. In particular, the Purcell factor reaches the maximum with $\phi\sim3\pi/4$, accompanied by an eightfold enhancement and order-of-magnitude linewidth narrowing compared to that of hybrid cavity. This is explained by the complex constructive interference of cavity modes at CEP. It should be emphasized that in hybrid CEP cavity, the lineshape of Purcell factor is no longer Lorentzian, in stark contrast to the hybrid cavity where the Purcell factor can still be well approximated by Lorentzian function in the case of far red-detuned plasmon-photon coupling, which is a special case of Fano resonance with a large Fano parameter $q$ \cite{RN14,RN18,RN57}. 

Distinguished from the low quality factor ($Q$) of plasmonic antenna, which is $Q_a=\omega_a/\kappa_a=10\sim20$ for common structures \cite{RN1,RN2,RN4,RN16}, the WGM cavity covers a wide range of $Q$ varying from several hundred to million \cite{RN52,RN54,RN58,RN59}, according to the different materials and structure geometries. Therefore, Figs. \ref{fig3}(a) and (b) investigate the Purcell enhancement $F_p/F_p^0$ and the corresponding linewidth narrowing $\Gamma_{EP}/\Gamma_{EP}^0$ of hybrid CEP cavity versus the quality factor $Q_c=\omega_c/\kappa$ of WGM modes and $g_1$, respectively, where $F_p=\text{max}[J(\omega)/J_0 (\omega)]$ is the maximal Purcell factor and $\Gamma_{EP}$ is the linewidth of hybrid CEP cavity, while $F_p^0$ and $\Gamma_{EP}^0$ denote the corresponding quantities of hybrid cavity. The results exhibit two remarkable features. One is that the Purcell enhancement can be always achieved in the chosen parameter ranges, as Fig. \ref{fig3}(a) shows, and a maximum of eightfold enhancement can be obtained for arbitrary $Q_c$ with an optimal $g_1$, which is denoted as $g_1^\text{opt}$ and indicated by the white dashed line in Figs. \ref{fig3}(a) and (b). However, a smaller $Q_c$ requires stronger plasmon-photon interaction to reach an eightfold Purcell enhancement. On the other hand, Fig. \ref{fig3}(b) shows that $g_1^\text{opt}$ is exactly corresponding to the greatest linewidth narrowing, where we find $\Gamma_{EP}/\Gamma_{EP}^0\sim0.1$ and is almost unchanged as $Q_c$ varies. From the analytical expression of spectral density [Eq. (\ref{eq11})], we find $g_1^{\text{opt}}\approx-\sqrt{3\Delta_{ac} \kappa_c }/2$ and the maximal Purcell enhancement (also the greatest linewidth narrowing) is achieved at $\omega\approx-3\kappa_c/2$, where $\Delta_{ac}=\omega_a-\omega_c$ is the plasmon-photon detuning. In Fig. \ref{fig3}(c), we compare the analytically predicted $g_1^\text{opt}$ (solid line) with the numerical results (hollow circles), where good agreement can be seen. It shows that WGM cavity with a moderate quality factor of $Q_c=10^3\sim10^4$ is suitable to demonstrate the enhanced Purcell effect at CEP since it yields $\left|g_1^\text{{opt}} \right|=10\sim30\text{meV}$, which is attainable in common plasmonic-photonic cavities.

\section{Enhanced coherent light-matter interaction at CEP}\label{sec4}

The Purcell enhancement and linewidth narrowing at CEP is expected to enhance the coherent light-matter interaction. This can be revealed from the emission spectrum of QE, which is defined as $S(\omega)=\lim _{t \rightarrow \infty} \operatorname{Re}\left[\int_0^{\infty} d \tau e^{i \omega \tau} \left\langle\sigma_{+}(t) \sigma_{-}(t+\tau)\right\rangle \right]$ \cite{RN50}, where the two-time correlation function $\left\langle\sigma_{+}(t) \sigma_{-}(t+\tau)\right\rangle$ can be calculated using the quantum regression theorem in a fashion similar to Eqs. (\ref{eq7})-(\ref{eq10}). The emission spectrum of QE is expressed as (see Appendix \ref{ac} for detailed derivation) 
\begin{equation}\label{eq12}
S(\omega)=\frac{1}{\pi} \frac{\gamma+\Gamma(\omega)}{\left[\omega-\omega_0-\Delta(\omega)\right]^2+\left[\frac{\gamma+\Gamma(\omega)}{2}\right]^2}
\end{equation}
where the local coupling strength $\Gamma(\omega)$ and the photon-induced Lamb shift $\Delta(\omega)$ are related to the cavity polarizability and given by $\Gamma(\omega)=-2g_a^2 \operatorname{Im}[\chi_{sys}(\omega)]$ and $\Delta(\omega)=g_a^2 \operatorname{Re}[\chi_{sys}(\omega)]$, respectively. The temporal dynamics of QE can be retrieved from the Fourier transform of the emission spectrum.
\begin{figure*}[t]
\centering\includegraphics[width=0.94\linewidth]{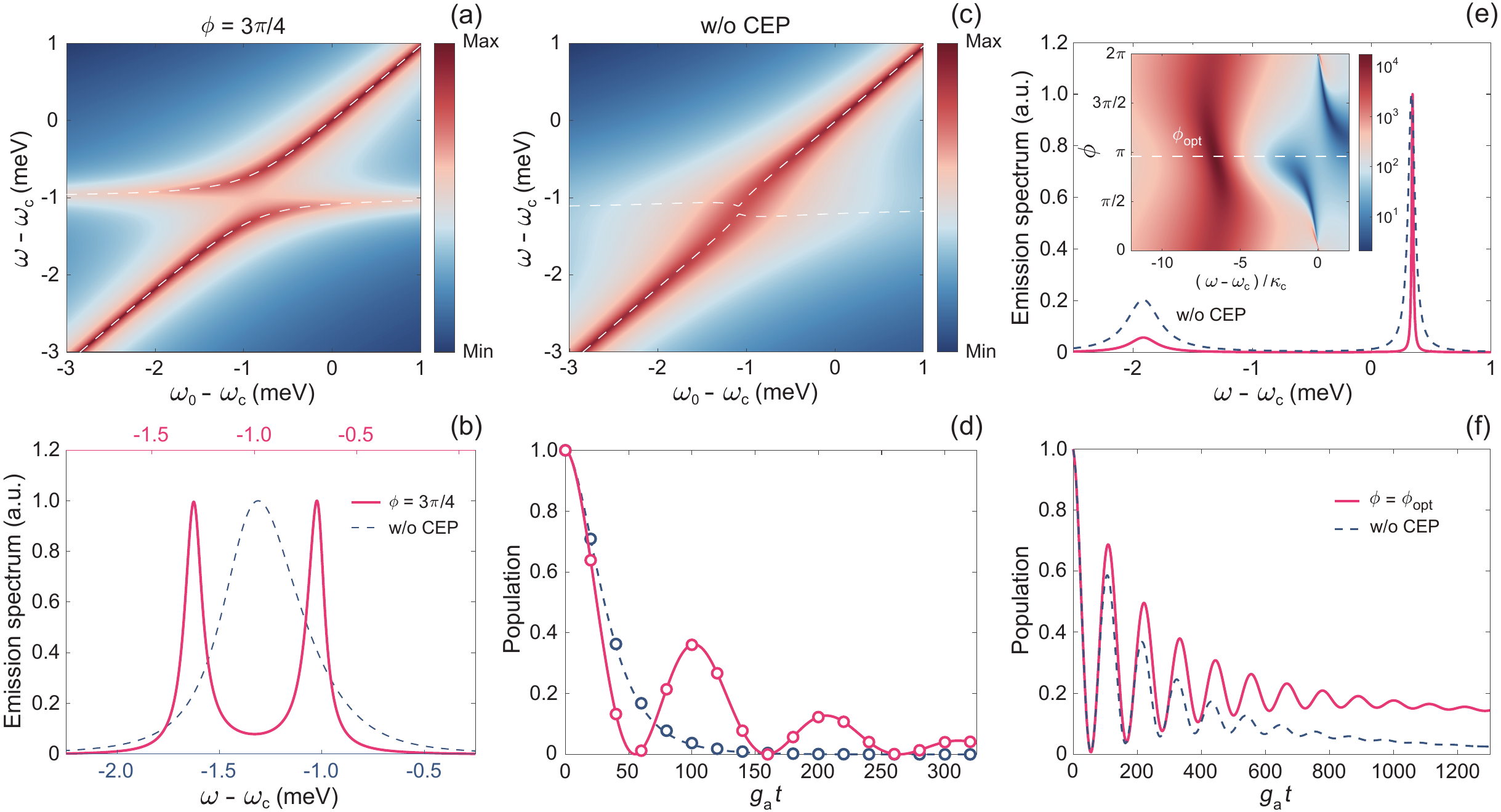}
\caption{CEP-enhanced quantum coherence. (a) and (c) Emission spectrum of QE versus QE-cavity detuning for hybrid CEP cavity with $\phi=3\pi/4$ and for hybrid cavity without CEP, respectively. The white dashed lines trace the eigenenergies of $\mathbf{M}_0$ [Eq. (\ref{eq4})]. (b) and (d) Emission spectrum of QE with equal splitting and the corresponding temporal dynamics, respectively. The emission spectrum of QE in hybrid CEP cavity in (b) is shifted to the peak location of that without CEP for the sake of comparison, and the corresponding frequency is given in the upper $x$-axis. The circles in (d) plot the results obtained by Fourier transforming the emission spectrum of QE, while the solid lines show the results obtained by numerically calculating the extended cascaded quantum master equation [Eqs. (\ref{eq1}) and (\ref{eq2})] using QuTip \cite{RN60}. The parameters are $\gamma_0=3\mu\text{eV}$, $g_1=-24\text{meV}$ and $Q_c=10^3$. (e) Comparison of the emission spectrum of QE in the strong-coupling regime. The pink solid line indicates the emission spectrum of QE in hybrid CEP cavity with $\phi=\phi^{\text{opt}}$ corresponding to the maximal Purcell factor, while the blue dashed line shows the emission spectrum of QE in hybrid cavity without CEP. The inset shows the Purcell factor of hybrid CEP cavity as the function of frequency and $\phi$, where the horizontal dashed line indicates $\phi^{\text{opt}}$. The parameters are $g_1=-20\text{meV}$, $g_a=40\text{meV}$ and $Q_c=10^4$. (f) shows the corresponding temporal dynamics of QE. The results are obtained by Fourier transforming the emission spectrum. Parameters not mentioned are the same as Fig. \ref{fig2}. }
\label{fig4}
\end{figure*}

Fig. \ref{fig4}(a) plots the emission spectrum of QE corresponding to the maximal Purcell enhancement in hybrid CEP cavity (i.e., $\phi=3\pi/4$), where the strong-coupling anticrossing can be clearly seen by varying the transition frequency of QE. While without CEP, no Rabi splitting can be seen in the emission spectrum of QE, as Fig. \ref{fig4}(c) shows. By comparing with the eigenenergies (dashed lines), we can see that the peak location of emission spectrum exactly corresponds to the transition frequency of QE, with broaden linewidth around cavity resonance due to the Purcell effect. This reflects the fact that the QE-cavity interaction is under the weak-coupling regime. Moreover, the strong light-matter interaction occurred at hybrid CEP cavity is also evident by the well-separated Rabi splitting in the emission spectrum of QE [Fig. \ref{fig4}(b)] and the prominent Rabi oscillation in temporal domain [Fig. \ref{fig4}(d)]. 

It is worth noting that the Rabi oscillation decaying with a rate $\Gamma_p$ smaller than the emission rate $\gamma_{\text{eff}}$ of QE in the weak-coupling regime, as we see in Fig. \ref{fig4}(d), is a counterintuitive behavior. It is because in the standard cavity QED system (i.e., the Jaynes–Cummings model) \cite{RN50}, we have $\Gamma_p=\kappa_c/2$ and $\gamma_{\text{eff}}=4g^2/\kappa_c$ when $\gamma_0 \ll \kappa_c$, where $g$ is the coupling rate of QE-cavity interaction. On the other hand, the critical coupling rate of the onset of strong coupling is $g_0=\kappa_c/4$ \cite{RN61}, which yields $\Gamma_p>\gamma_{\text{eff}}$, i.e., the Rabi oscillation should decay faster than a QE weakly coupled to the cavity. Therefore, the results of Fig. \ref{fig4}(d) implies that the composed system enters into the strong-coupling regime mainly attributed to the effect of linewidth narrowing at CEP. This feature is more evident for a composed system where the QE-cavity interaction is already in the strong-coupling regime without CEP, see Fig. \ref{fig4}(e) for an example. In such a case, we find the locations of two Rabi peaks in the emission spectrum of QE remain approximately unchanged in the presence of CEP, so does the period of Rabi oscillation, as shown in Fig. \ref{fig4}(f). This indicates that the effective coupling rate between the QE and cavity is similar to the case without CEP. However, the linewidth of Rabi peaks is significantly narrow and as a consequence, the Rabi oscillation manifests a slower decay. The results demonstrate the enhanced quantum coherence by CEP in both the weak- and strong-coupling regimes.

\section{Enhanced quantum yield at CEP}\label{sec5}

In the above section, we discuss the enhanced coherent light-matter interaction in hybrid CEP cavity with off-resonant plasmon-photon coupling. In the following, we demonstrate the possibility of enhancing the quantum yield by CEP in the case of resonant plasmon-photon coupling, where low quantum yield is expected due to the severe nonradiative loss of plasmonic resonance. The equations for evaluating the quantum yield are given by
\begin{equation}\label{eq13}
\frac{d}{d t} \vec{p_0}=-i \mathbf{M_0} \vec{p_0}-i \mathbf{\Omega}
\end{equation}
where the frequency in the diagonal elements of $\mathbf{M_0}$ is now replaced by the frequency detuning $\Delta_L=\omega_X-\omega_L$ between the system constituents and the driving field, with $X=0,a,ccw,cw$ and $\omega_L$ being the frequency of driving field. The composed system is initially in the ground state and $\mathbf{\Omega}=\left[p_{in},0,0,0\right]^T$ accounts for a weak coherent drive of QE with amplitude $p_{in}$, which is introduced by implementing a driving Hamiltonian $H_d=p_{in} \left(e^{-i\omega_L t} \sigma_{+}+\sigma_{-} e^{i\omega_L t} \right)$ in Eq. (\ref{eq1}). The steady-state solutions of Eq. (\ref{eq12}) are used to calculate the quantum yield, which is defined as $\eta=\Phi_r/\left(\Phi_r+\Phi_d \right)$, with radiation power $\Phi_r=\langle\left(\sqrt{\kappa_r} a^{\dagger}+\sqrt{\gamma_0} \sigma_{+} \right)\left(\sqrt{\kappa_r} a+\sqrt{\gamma_0} \sigma_{-} \right)\rangle+\kappa_c \left(\langle c_{ccw}^{\dagger} c_{ccw} \rangle+\langle c_{cw}^{\dagger} c_cw \rangle \right)$ and absorption power of plasmonic modes $\Phi_d=\kappa_o \langle a^{\dagger} a \rangle+\gamma_m \langle \sigma_{+} \sigma_{-} \rangle$, where $\gamma_m$ accounts for the dissipation of QE to higher-order plasmonic modes, i.e., the QE decay rate becomes $\gamma=\gamma_0+\gamma_{nr}+\gamma_m$ in Eq. (\ref{eq2}). The quantum yield, radiation and absorption power of hybrid cavity are denoted as $\eta_0$, $\Phi_r^0$ and $\Phi_d^0$, respectively. It is worth mentioning that the first term in $\Phi_r$ stands for the radiation from both plasmonic antenna and QE, considering that the differences of emission from these two constituents to detector can be neglected due to their subwavelength dimensions.  

\begin{figure}[t]
\centering\includegraphics[width=0.92\linewidth]{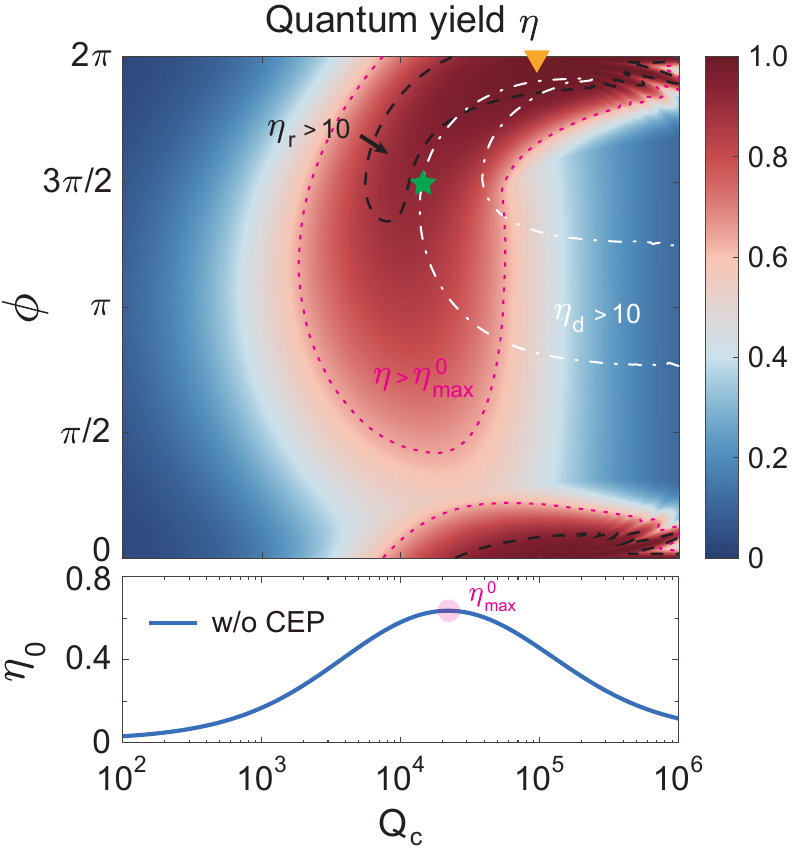}
\caption{Quantum yield $\eta$ of hybrid CEP cavity versus $Q_c$ and $\phi$. The red dotted line surrounds the region of enhanced quantum yield at CEP, i.e., $\eta>\eta_0$. The black dashed line shows the parameter area of cavity radiation enhancement $\eta_r=\operatorname{max}\left[\Phi_r \right]/\operatorname{max}\left[\Phi_r^0 \right]>10$, while the white dashed dotted line indicates the parameter area of plasmon absorption reduction $\eta_d=\operatorname{min}\left[\Phi_d^0 \right]/\operatorname{min}\left[\Phi_d \right]>10$ for dipolar plasmonic mode. The yellow triangle and green star indicate the parameters of Figs. \ref{fig6}(a) and (b), respectively. The bottom panel shows the quantum yield $\eta_0$ of hybrid cavity without CEP. Other parameters are the same as Ref. \cite{RN15}: $g_c=0.144\text{meV}$, $g_a=7.2\text{meV}$, $g_1=-2.9\text{meV}$, $\kappa_r=2.45\text{meV}$, $\kappa_o=200\text{meV}$, $Q_c=10^5$, $\gamma_0=3\mu \text{eV}$, $\gamma_m=83\mu \text{eV}$ and $\gamma_{nr}=0$. }
\label{fig5}
\end{figure}

To better illustrate the enhancement of quantum yield at CEP, we first adopt the parameters of a hybrid cavity reported in Ref. \cite{RN15} at low temperature. The results are presented in Fig. \ref{fig5}, where the main panel shows the quantum yield $\eta$ with CEP versus $Q_c$ and $\phi$; meanwhile, the results without CEP is provided in the bottom panel for comparison. We can see that the maximal quantum yield without CEP is $\eta_{\text{max}}^0=0.636$, achieved at $Q_c\sim2\times10^4$; while in a hybrid CEP cavity, the enhanced quantum yield (i.e., $\eta>\eta_{\text{max}}^0$) can be found with $Q_c$ varying from $2\times10^3$ to $10^6$, indicated by the red dashed dotted line. In addition, high quantum yield with $\eta>0.8$ can be achieved in a wide range of parameters. 

\begin{figure}[t]
\centering\includegraphics[width=0.98\linewidth]{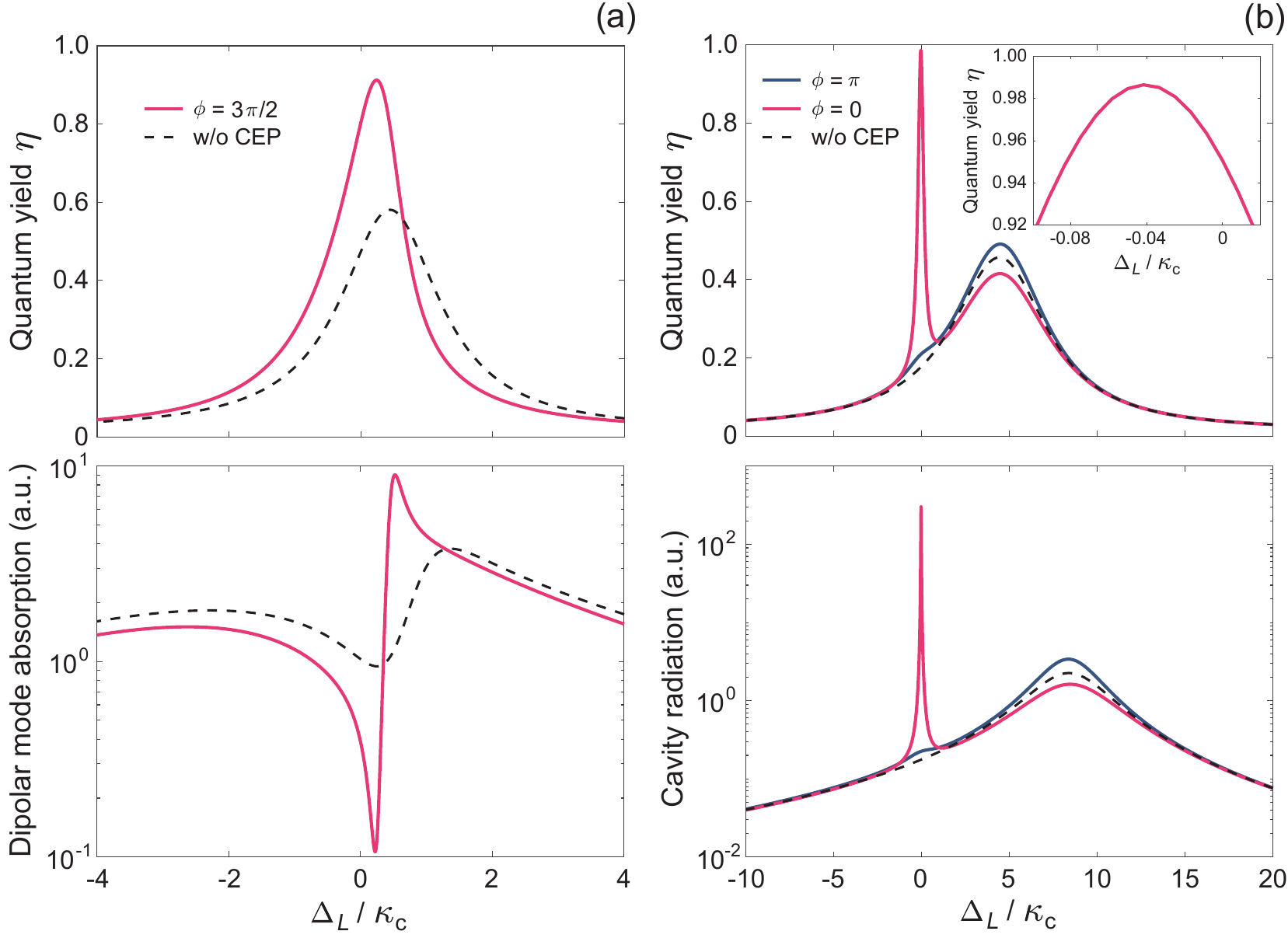}
\caption{Reduced plasmonic absorption and enhanced cavity radiation for high quantum yield of hybrid CEP cavity. (a) Quantum yield (upper panel) and plasmon absorption of dipolar mode (lower panel) in hybrid CEP cavity for $Q_c\sim1.5\times10^4$ (see the green star in Fig. \ref{fig5}). (b) Quantum yield (upper panel) and cavity radiation (lower panel) of hybrid CEP cavity for $Q_c=10^5$ (see the yellow triangle in Fig. \ref{fig5}). The inset shows a close-up of quantum yield around $\Delta_L=0$. In (a) and (b), the results of hybrid cavity without CEP are also shown for comparison (black dashed lines). Parameters not mentioned are the same as Fig. \ref{fig5}. }
\label{fig6}
\end{figure}

To shed insight on the physical mechanism of enhanced quantum yield at CEP, we plot $\eta$ as the function of frequency detuning $\Delta_L$ in the upper panels of Fig. \ref{fig6}, where two situations are analyzed. The parameters of the first are similar to hybrid cavity achieving $\eta_{\text{max}}^0$, with $Q_c\sim1.5\times10^4$ and $\phi=3\pi/2$, as the green star in Fig. \ref{fig5} indicates. The maximal quantum yield $\eta_{\text{max}}$ of hybrid CEP cavity is enhanced by $50\%$ and can reach 0.92, as we see in the upper panel of Fig. \ref{fig6}(a). This enhancement stems from the CEP-reduced absorption of dipolar plasmonic mode, see the absorption power $\Phi_d$ shown in the lower panel of Fig. \ref{fig6}(a). It shows that the minimum of absorption power is reduced by one order of magnitude in the presence of CEP, and the valley of absorption power is exactly corresponding to $\eta_{\text{max}}$. This feature indicates that the enhanced quantum yield originates from the Fano interference, as found by the authors in Ref. \cite{RN15}, while in our setup this mechanism is strengthened by CEP. The white dotted line in Fig. \ref{fig5} surrounds the region of $\eta_d=\operatorname{min}\left[\Phi_d^0 \right]/\operatorname{min}\left[\Phi_d \right]>10$, indicating the parameter area of significantly reduced plasmonic absorption. We can see that for hybrid CEP cavity with high $Q_c$, a high quantum yield is still hard to achieve by means of reducing the plasmonic absorption. In such a case, the improvement of quantum yield is hindered by the low radiation power of cavity. 

We then focus on the region of high quantum yield extended to high $Q_c$ at the upper right corner of Fig. \ref{fig5}, with $\phi$ around zero (or 2$\pi$). The quantum yield $\eta$ versus $\Delta_L$ for $\phi=0$ and $Q_c=10^5$ (indicated by the yellow triangle in Fig. \ref{fig5}) is analyzed in the upper panel of Fig. \ref{fig6}(b). It shows that high quantum yield can be achieved in a narrow range of frequency around $\Delta_L=0$, where $\eta$ can reach 0.98 while the maximal quantum yield without CEP ($\eta^0_{\text{max}}$) is below 0.5. We notice that in hybrid CEP cavity, the frequency corresponding to the maximal quantum yield $\eta_{\text{max}}$ deviates from that of hybrid cavity; actually, the quantum yield with $\phi=0$ is slightly reduced at the frequency of $\eta_{\text{max}}^0$. While for $\phi=\pi$, which corresponds to the region of Fano-reduced plasmonic absorption, only a weak enhancement of quantum yield is observed at the frequencies of $\eta_{\text{max}}$ and $\eta_{\text{max}}^0$. Therefore, the results clearly indicate that the underlying mechanism of $\eta_{\text{max}}$ achieved at CEP is essentially different from the Fano-reduced plasmonic absorption of hybrid cavity. By inspecting the output power of cavity radiation $\Phi_r$ [the lower panel of Fig. \ref{fig6}(b)], we find a sharp peak appearing exactly at the frequency of $\eta_{\text{max}}$, with an enhancement of about three orders of magnitude compared to that of hybrid cavity. It reveals that the striking improvement of quantum yield with $\phi=0$ comes from the enhanced cavity radiation at CEP, which can be interpreted as a phenomenon analogous to superscattering where the partial radiation of eigenmodes is added up to produce an extremely sharp peak in $\Phi_r$ without interference \cite{RN63}. To the best of our knowledge, this mechanism of enhaned quantum yield, i.e., the superscattering at CEP, has not been reported in plasmonic-photonic cavity. The black dashed line in Fig. \ref{fig5} indicates the region of $\eta_r=\operatorname{max}\left[\Phi_r \right]/\operatorname{max}\left[\Phi_r^0 \right]>10$, where we see that the prominent superscattering enables high quantum yield with high $Q_c$ around $\phi=0$ (or $2\pi$). 

To gain more insight on how the cavity radiation is enhanced by the superscattering at CEP, we derive a formalism of eigenmode decomposition for the scattering spectrum of hybrid CEP cavity \cite{RN63}. In absence of QE, the dynamic of cavity modes is described by $d\langle \vec{c}(t) \rangle/dt=-i\mathbf{M_c} \langle \vec{c}(t) \rangle - i \mathbf{\Omega_p}$, where the frequency in the diagonal elements of $\mathbf{M_c}$ is again replaced by the frequency detuning $\Delta_L$, and $\mathbf{\Omega_p}=\left[q_{in},0,0\right]^T$ accounts for the driving field for dipolar plasmonic mode. This equation can be rewritten as 
\begin{equation}\label{eq14}
i \frac{d}{d t}\langle\vec{c}(t)\rangle=V^{-1} B V\langle\vec{c}(t)\rangle+\boldsymbol{\Omega}_p
\end{equation}
where $B$ and $V$ are the diagonal matrix formed from the eigenvalues of $\mathbf{M_c}$ and the corresponding left eigenvectors in matrix form, respectively. The above equation can be solved through the Fourier transform. The solution is
\begin{equation}\label{eq15}
\vec{c}\left(\Delta_L\right)=V^{-1}\left(\Delta_L I-B\right)^{-1} V \mathbf{\Omega}_p
\end{equation}
The scattering of hybrid CEP cavity is expressed as $\sigma(\Delta_L )=s^{\dagger} (\Delta_L )s(\Delta_L)$, with $s(\Delta_L)=K \vec{c}(\Delta_L)$, where $K$ defines the coupling $\Gamma$ between different scattering channels, which are independent in our setup
\begin{equation}\label{eq16}
\Gamma=K^{\dagger} K=\left[\begin{array}{ccc}
\kappa_a & 0 & 0 \\
0 & \kappa_c & 0 \\
0 & 0 & \kappa_c
\end{array}\right]
\end{equation}
After lengthy calculations (see Appendix \ref{ad} for details), we can obtain the expression of $\sigma(\Delta_L)$ at $\Delta_L=0$
\begin{equation}\label{eq17}
\begin{gathered}
\sigma_0 \equiv \sigma(0)=\frac{p}{\gamma_a \gamma_b \gamma_c}\left\{h_{a a}\left|C_a\right|^2+h_{b b}\left|C_b\right|^2+h_{c c}\left|C_c\right|^2\right. \\
\left.+2 \operatorname{Re}\left[h_{a b} C_a^* C_b+h_{a c} C_a^* C_c+h_{b c} C_b^* C_c\right]\right\}
\end{gathered}
\end{equation}
where $p=\operatorname{Det}\left[V\right]^{-2}$. $C_i$ is defined as the radiation pattern of eigenmodes, with $i=a,b,c$, while $h_{ij}$ and $h_{ii}$ encode the interaction and the weighted coefficient of eigenmodes, respectively. For the specific expressions of $C_i$, $h_{ii}$ and $h_{ij}$, we refer to Appendix \ref{ad}. $\gamma_i$ is the imaginary part of eigenvalues of $\mathbf{M_c}$ in decreasing order, i.e., $\gamma_a>\gamma_b>\gamma_c$. Therefore, eigenmode $a$ is superradiant. We denote the first term as $\sigma_{\text{sup}}$ while the remaining terms as $\sigma_{\text{so}}$, thus $\sigma_0=\sigma_{\text{sup}}+\sigma_{\text{so}}$. Therefore, if the condition $\gamma_a \gg \gamma_b,\gamma_c$ is satisfied, which holds true for hybrid CEP cavity, the superscattering will occur when $\sigma_{\text{sup}}>\sigma_{\text{so}}>0$, giving rise to a peak in the scattering spectrum.

\begin{figure*}[t]
\centering\includegraphics[width=0.97\linewidth]{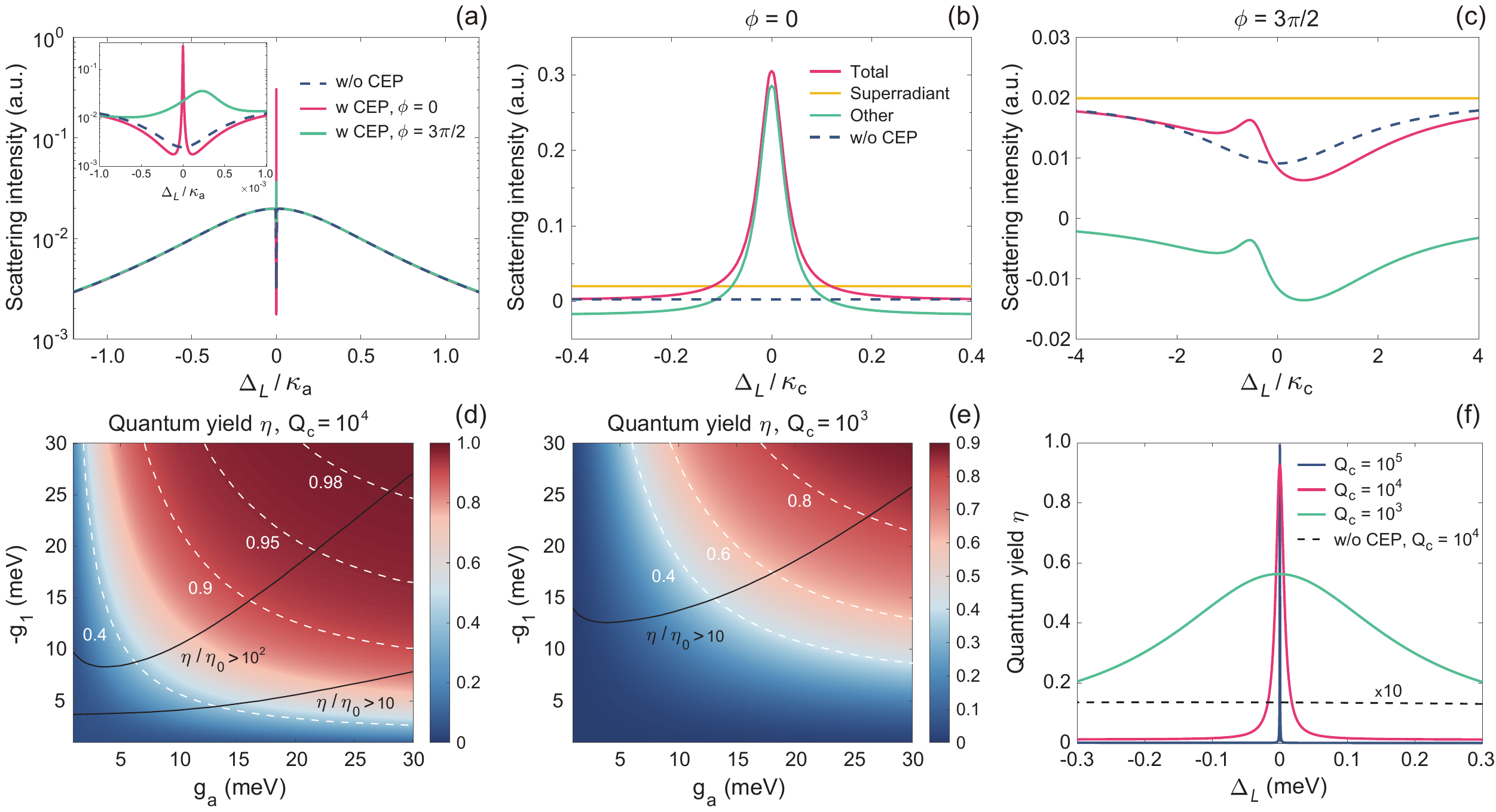}
\caption{Superscattering and near-unity quantum yield at room temperature. (a) Scattering spectrum $\sigma(\Delta_L)$ of hybrid CEP cavity for plasmon driving. The inset shows a close-up of $\sigma(\Delta_L)$ around $\Delta_L=0$. (b) and (c) Decomposition of cavity scattering $\sigma(\Delta_L)$ into the superscattering ($\sigma_{\text{sup}}$) and other ($\sigma_{\text{so}}$) terms for the cases of $\phi=0$ and $3\pi/2$, respectively. Parameters of these two cases are indicated by the yellow triangle and green star in Fig. \ref{fig6}, respectively. The results of hybrid cavity without CEP are also shown for comparison (dashed lines). (d) and (e) Room-temperature quantum yield $\eta$ of hybrid CEP cavity versus $g_a$ and $g_1$ for $Q_c=10^4$ and $10^3$, respectively. $\eta_0$ denotes the quantum yield of hybrid cavity without CEP. (f) Comparison of room-temperature quantum yield versus the frequency detuning for various $Q_c$. The quantum yield of hybrid cavity without CEP is also shown for comparison (black dashed line). The parameters are $\phi=0$, $g_1=-15$meV, $g_a=20$meV and $\gamma_{nr}=15$meV. Other parameters are the same as Fig. \ref{fig6}. }
\label{fig7}
\end{figure*}

Fig. \ref{fig7}(a) compares $\sigma(\Delta_L)$ with $\phi=0$ and $3\pi/2$, while other parameters are the same as Fig. \ref{fig6}(b). We can see that a sharp peak shows around $\Delta_L=0$ for both cases, but the intensity of $\phi=0$ is much higher than that of $\phi=3\pi/2$. On the contrary, no peak can be found around $\Delta_L=0$ without CEP. In Figs. \ref{fig7}(b) and (c), we plot the corresponding eigenmode decomposition of $\sigma(\Delta_L)$ around $\Delta_L=0$. We find $\sigma_{\text{so}}\gg\sigma_{\text{sup}}$ for $\phi=0$; furthermore, the scattering peak is mainly contributed by $\sigma_{\text{so}}$ and demonstrates a subnatural linewidth. These features signify the occurrence of superscattering at CEP. While for $\phi=3\pi/2$, Fig. \ref{fig7}(c) shows a scattering peak appearing at the local maximum of $\sigma_{\text{so}}$, however, in this case $\sigma_{\text{so}}$ is negative around $\Delta_L=0$, i.e., $\sigma_{\text{sup}}>0>\sigma_{\text{so}}$. Therefore, it belongs to the intermediate mechanism of electromagnetically induced transparency and superscattering.

In view of the advance of hybrid CEP cavity in enhancing quantum yield at low temperature, we investigate the performance at room temperature. In this case, the intrinsic quantum yield of QE [i.e., $\gamma_0/(\gamma_0+\gamma_m+\gamma_{nr})$] approaches to zero, while the quantum yield of hybrid cavity is dramatically reduced by the nonradiative decay $\gamma_{nr}$ of QE. We evaluate that the quantum yield of hybrid cavity decreases from 0.58 to less than 0.014 for $Q_c=10^4$ and $\gamma_{nr}=15$meV \cite{RN64,RN65}. With the same $Q_c$ and $\gamma_{nr}$, Fig. \ref{fig7}(d) displays the quantum yield of hybrid CEP cavity as the function of $g_a$ and $g_1$, where it shows that $\eta>0.9$ can be achieved with $g_a$,$\left|g_1\right|>15$meV. It also shows that strong plasmon-photon and plasmon-QE interactions are beneficial to improve quantum yield. $\eta$ rapidly grows from $\sim0.4$ to $\sim0.9$ as the coupling rates $g_a$ and $\left|g_1\right|$ increase from several meV to 15meV. With $g_a$,$\left|g_1\right|\sim30$meV, which are attainable in realistic structures \cite{RN25,RN48}, near-unity quantum yield can be achieved at room temperature, demonstrating over hundredfold enhancement of quantum yield. As $Q_c$ reduces to $10^3$, Fig. \ref{fig7}(e) shows that the quantum yield significantly decreases in a wide range of parameters, but hybrid CEP cavity with stronger plasmon-photon and plasmon-QE interactions manifests higher robustness. For example, the quantum yield drops from 0.88 to 0.45 for $g_a$,$\left|g_1 \right|=15$meV, while high quantum yield $\eta_{\text{max}}\approx0.9$ can still be maintained with $g_a$,$\left|g_1\right|=30$meV. 

The results presented in Figs. \ref{fig7}(d) and (e) indicate that a high $Q_c$ facilitates to improve the quantum yield of hybrid CEP cavity. Fig. \ref{fig7}(f) investigates the quantum yield for various $Q_c$ versus $\Delta_L$, where $\eta_{\text{max}}$ is found to reach 0.992 at room temperature with $Q_c=10^5$. It also indicates that this near-unity quantum yield is achieved by the superscattering at CEP, while the mechanism of Fano interference fails to produce high quantum yield at room temperature. Therefore, the results demonstrate the unique advance of hybrid CEP cavity in realizing high quantum yield and great potential in practical applications, such as building room-temperature single-photon sources \cite{RN66,RN67}.

\section{A physical realization}\label{sec6}

\begin{figure*}[t]
\centering\includegraphics[width=0.92\linewidth]{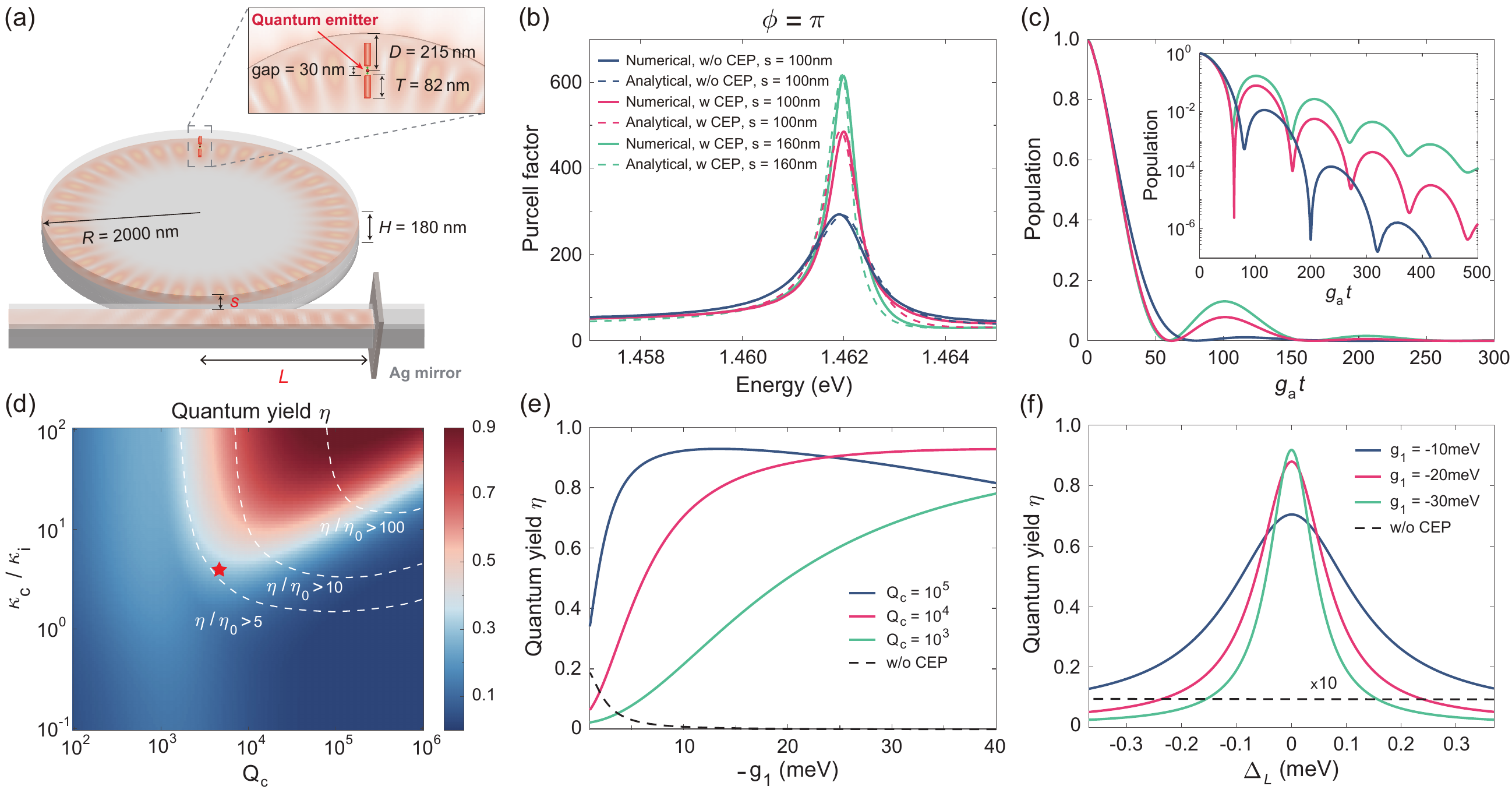}
\caption{Enhanced coherent light-matter interaction and room-temperature quantum yield of a realistic hybrid CEP cavity. (a) Schematic diagram of a hybrid CEP cavity based on WGM microdisk. A gold dimer consisting of two nanorods is placed on the top of microdisk, with a QE located in the gap center and aligned matching the polarization of dipolar plasmonic mode. The radius of nanorods is 40nm. Other geometry parameters are indicated in the figure. The Purcell factor can by tailored by adjusting the edge-to-edge separation $s$ between the cavity and the waveguide and the distance $L$ between the cavity-waveguide junction and the mirror made of Ag \cite{RNag}. The width of waveguide is 400nm. The orange color illustrates the field profiles for $\phi=\pi$ and $\omega_c=1.462$eV. (b) Purcell factor of the physical realization shown in (a) for various $s$ obtained from electromagnetic simulations. The dashed lines with the same color plot the corresponding analytical results given by the quantized few-mode model [Eqs. (\ref{eqb11})]. (c) Temporal dynamics of initially excited QE with a dipole moment $\mu=48$ Debye for various Purcell factors shown in (b). The inset presents the logarithmic plot of QE population. (d) Room-temperature quantum yield $\eta$ of hybrid CEP cavity as the function of $Q_c$ and $\kappa_c/\kappa_i$. $\eta_0$ denotes the quantum yield of hybrid cavity without CEP. The red star indicates $Q_c$ and $\kappa_c/\kappa_i$ of WGM modes with frequency $\omega_c=2.246$eV. Other parameters are $g_1=-11$meV, $g_a=23.6$meV, $g_c=0.171$meV and $\gamma_{nr}=15$meV. (e) Room-temperature quantum yield $\eta$ versus $g_1$ for various $Q_c$. The frequencies and decay rates of uncoupled gold dimer and WGM cavity are provided in the text. (f) Room-temperature quantum yield $\eta$ for various $g_1$. The parameters are $\kappa=\omega_c/10^4$ and $\kappa_c=10^2 \kappa_i$. Other parameters are the same as (d). }
\label{fig8}
\end{figure*}

In the above discussion, we omit the intrinsic decay of WGM modes. We will address this issue in this subsection by evaluating the quantum yield of QE in a realistic structure of hybrid CEP cavity. To do this, we need to extract the system parameters from the simulation data by applying the analytical LDOS. Therefore, we should first validate our LDOS theory.

Fig. \ref{fig8}(a) depicts the geometry of hybrid CEP cavity under study, which is based on a SiN microdisk with a gold dimer placed on the top surface. The refractive index of microdisk is $n=2$, with an imaginary component of $4\times10^{-6}$ to include the material absorption \cite{RN14}. The permittivity $\epsilon(\omega)$ of gold is characterized by the Drude model $\epsilon(\omega)=1-\omega_g^2/\omega(\omega+i\gamma_g)$, where $\omega_g=1.26\times10^{16}$ rad/s and $\gamma_g=1.41\times10^{14}$ rad/s are the plasma frequency and the collision rate, respectively \cite{RN23}. The gold dimer on SiN slab supports a single localized plasmon resonance with the peak appearing at $\omega_a=2.254$eV and a linewidth (total decay rate) of $\kappa_a=255$meV, where the radiative decay rate is evaluated as $\kappa_r=13$meV. A point-dipole QE is located at the gap center of the gold dimer.

According to the LDOS theory developed in Sec. \ref{sec2}, the Purcell enhancement felt by QE can be modified by adjusting the cavity-waveguide separation $s$ and the distance $L$ between the cavity-waveguide junction and the mirror, as the former determines $\kappa_c$ while the latter controls $\phi$. To verify our LDOS theory, we consider a pair of degenerate WGM modes with frequency $\omega_c=1.462$eV and intrinsic linewidth $\kappa_i=0.387$meV in absence of waveguide. The frequency of cavity resonance is unaffected by the coupling to waveguide with $s=100$nm, but the linewith is broadened to $\kappa$, from which we obtain $\kappa_c=0.867$meV by subtracting the intrinsic linewidth from $\kappa$. With the parameters of uncoupled constituents at hand, we can employ the analytical expression of complete spectral density [i.e., including $J_c(\omega)$ and $J_{ac}(\omega)$] given in Eqs. (\ref{eqb11}) to determine the plasmon-photon coupling rate $g_1$ by fitting the Purcell factor obtained from electromagnetic simulations. We evaluate $g_1=-11$meV and find good agreement between the analytical and simulation results, see the blue dashed and solid lines in Fig. \ref{fig8}(b). Then we introduce the mirror at the right end of waveguide to create a CEP with $\phi=\pi$, which corresponds to the maximal Purcell enhancement of this setup. We find the analytically predicted Purcell factor accords well with the simulation results, see the pink solid and dashed lines in Fig. \ref{fig8}(b). In addition, for large cavity-waveguide separation $s=160$nm, the analytical model can also give a correct prediction of Purcell factor. Therefore, the results of Fig. \ref{fig8}(b) validate the applicability of our LDOS theory for hybrid CEP cavity. 

Fig. \ref{fig8}(b) shows that the Purcell factor is increased by $65\%$ after introducing a CEP and is double as the cavity-waveguide separation enlarges to 160nm. In Fig. \ref{fig8}(c), we plot the corresponding temporal dynamics of initially excited QE, where it shows the greater and faster Rabi oscillation with slower decay compared to that without CEP. This indicates the stronger coherent energy exchange between the cavity and QE as a result of the enhanced Purcell factor with slightly narrow linewidth at CEP. 

The intrinsic decay of cavity modes is inevitable in realistic structures, which hinders the improvement of quantum yield. Fig. \ref{fig8}(d) displays the quantum yield of hybrid CEP cavity versus $Q_c$ and $\kappa_c/\kappa_i$ with $\phi=0$, where the QE is resonantly coupled to WGM modes with frequency $\omega_c=2.246$eV. The corresponding intrinsic and waveguide-induced decay rates for $s=40$nm are indicated by the red star in Fig. \ref{fig8}(d), which yields $\eta\sim0.36$. Fig. \ref{fig8}(d) shows that the enhancement of quantum yield requires an increase of $Q_c$ while simultaneously reducing the intrinsic decay of WGM modes. This can be achieved by using a WGM cavity with large radius \cite{RN39} and made of high refractive index \cite{RN48}. With the fixed $\kappa_c/\kappa_i=100$, Fig. \ref{fig8}(e) shows that the quantum yield for $\left|g_1 \right|>5$meV can be significantly enhanced as $Q_c$ varies from $10^3$ to $10^4$, and reach 0.9 as $\left|g_1 \right|$ increases to 20meV. However, a high $Q_c$ may also have negative impact on the quantum yield. When $\left|g_1 \right|>25$meV, the increase of $Q_c$ from $10^4$ to $10^5$ leads to the reduction of quantum yield. It is because the radiation through cavity is less efficient for a high $Q_c$, then a substantial portion of energy is absorbed by plasmon with a large $g_1$ since the intrinsic quantum yield of QE is extremely low. In addition, Fig. \ref{fig8}(f) shows that a large $g_1$ narrows the linewidth of radiation spectrum and may lead to the reduced quantum yield outside the cavity resonance, which is detrimental to realize the broadband enhancement of quantum yield. Therefore, the results show that $g_1$ also plays an important role in determining the quantum yield, but is not always in a positive manner.

Finally, we briefly summarize and discuss the findings from this simple design of hybrid CEP cavity. To achieve high quantum yield at room temperature, a hybrid CEP cavity with moderate $\left|g_1 \right|\sim20$meV, relatively high $Q_c\sim10^5$ and low $\kappa_i$ are desirable. We note that it is not difficult to find a realistic structure satisfying one or two of them, but is challenging to meet all of these requirements in WGM cavity. It is because a high $Q_c$ in general means weak energy leakage, resulting in a small overlap between the plasmonic and photonic modes when the plasmonic antenna is placed outside the cavity. This results in the tradeoff between $Q_c$ and $g_1$ in hybrid CEP cavity. To overcome this obstacle, the plasmonic antenna can be embedded in the cavity, with an analogues design of the structure studied in Ref. \cite{RN48}; however, in such a case the QE location cannot be feasibly controlled in experiments. A possible solution is to introduce a thin air slot at the antinode of the selected WGM modes \cite{RN68}, where the plasmonic antenna and QE can be placed with precise control. The presence of air slot in WGM microdisk will not degrade $Q_c$ since it has a physical volume as small as the plasmonic antenna, thus can be seen as a perturbation of WGM modes. Such hybrid CEP cavity design can fulfill the requirements of achieving high quantum yield at room-temperature, and can also be integrated with other on-chip optoelectronic devices. 

\section{Conclusion and outlook}\label{sec7}
In this work, we propose to engineer the plasmonic resonance by virtue of a CEP hosting in a WGM cavity and demonstrate the great tunability of LDOS provided by CEP. A LDOS theory is established to reveal the Purcell enhancement accompanied by order-of-magnitude linewidth narrowing at CEP, which results in the enhanced coherent light-matter interaction and the reduced dissipation of polaritonic states. Importantly, we identify a new mechanism of enhancing the quantum yield through the superscattering at CEP, which holds great promise for realizing a near-unity quantum yield at room temperature. A physical implement is utilized to validate our LDOS theory and analyze the possible factors that hinder the experimental demonstration of the predicted high quantum yield. One direction of future study is the optimization of structure design with moderate plasmon-photon interaction, high quality factor and low intrinsic decay. We believe that our work can provide insights on harnessing the non-Hermiticity of open quantum systems in quantum states control, which may benefit diverse quantum-optics applications.

\begin{acknowledgments}
This work is supported by the National Natural Science Foundation of China (Grant Nos. 62205061, 12274192, 11874438) and the Key Project of Natural Science Foundation of Henan (Grant No. 232300421141). 
\end{acknowledgments}

\appendix

\section{Derivation of the extended cascaded quantum master equation}\label{aa}

\begin{figure}[t]
\centering\includegraphics[width=0.98\linewidth]{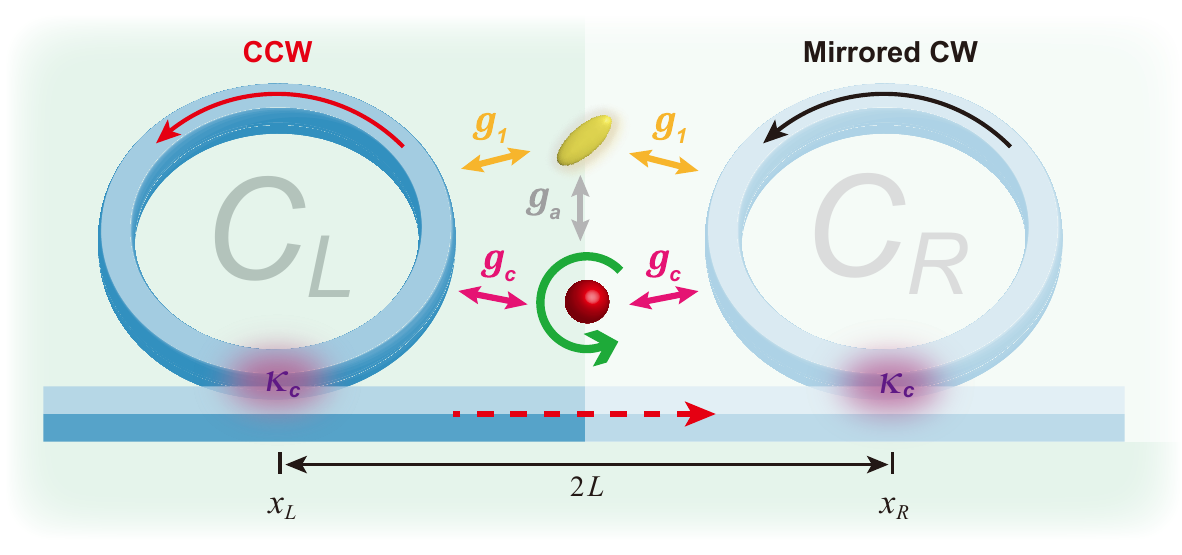}
\caption{Schematic diagram of the equivalent model of hybrid CEP cavity without the mirror. The original CW mode is flipped into a mirrored CCW mode and as a result, the QE becomes circularly polarized due to the mirror symmetry. $x_L$ and $x_R$ denote the positions of cavity modes. Other notions are the same as that in the main text. }
\label{figs1}
\end{figure}

This section is aimed to derive the quantum master equation (QME) for hybrid CEP cavity based on the theoretical framework of waveguide QED \cite{RNpra2015,RN30}, which is called the extended cascaded QME [Eqs. (\ref{eq1}) and (\ref{eq2})] since the CW mode is driven by the output field of CCW mode similar to a cascaded quantum system \cite{RN70}. We consider an equivalent model by removing the mirror, as depicted in Fig. \ref{figs1}, where the CW mode is flipped into a mirrored CCW mode through the mirror symmetry. Accordingly, the bosonic annihilation operators of the original and the mirrored CCW modes are denoted as $c_L$ and $c_R$, respectively. We can see that in this equivalent model, only the right-propagating guided mode of waveguide is involved in the interaction between $c_L$ and $c_R$. Then the extended cascaded QME can be derived from tracing out the waveguide modes. The system Hamiltonian including the waveguide modes is written as 
\begin{equation}
H_S=H+H_w+H_{s w}
\end{equation}
\noindent where $H=H_0+H_I$ is given in Eq. (\ref{eq1}). $H_w$ is the free Hamiltonian of waveguide
\begin{equation}
H_w=\int d \omega \omega b_R^{\dagger} b_R
\end{equation}
and $H_{sw}$ describes the Hamiltonian of cavity-waveguide interaction
\begin{equation}
H_{s w}=i \sum_{j=L, R} \int d \omega \sqrt{\frac{\kappa_c}{2 \pi}} b_R^{\dagger} e^{-i k x_j} c_j+H . c .
\end{equation}
where $b_R$ is the bosonic annihilation operator of the right-propagating waveguide mode with frequency $\omega$ and wave vector $k=\omega_c/v$ with $v$ being the group velocity. $x_{L}$ and $x_{R}$ are the positions of CCW mode and the mirrored CW mode, respectively. Without the loss of generality, we set $x_L=0$ and $x_R=2L$ according with the definition in the main text. Applying the transformation $\widetilde{H}=U H U^{\dagger}-i dU/dt U^{\dagger}$ with $U = \exp \left[i\left(\omega_c \sum_{j=L, R} c_j^{\dagger} c_j+\int d \omega \omega b_R^{\dagger} b_R\right)t\right]$, we have
\begin{equation}
\widetilde{H}_{s w}(t)=i \sum_{j=L, R} \int d \omega \sqrt{\frac{\kappa_c}{2 \pi}} b_R^{\dagger} e^{i\left(\omega-\omega_c\right) t} e^{-i \omega x_j / v} c_j+H . c .
\end{equation}
The equation of motion of $b_R$ can be obtained from the Heisenberg equation
\begin{equation}
\frac{d}{d t} b_R(t)=\sum_{j=L, R} \sqrt{\frac{\kappa_c}{2 \pi}} c_j e^{i\left(\omega-\omega_c\right) t} e^{-i \omega x_j / v}
\end{equation}
The above equation can be formally integrated to obtain
\begin{equation}
b_R(t)=\sum_{j=L, R} \int_0^t d \tau \sqrt{\frac{\kappa_c}{2 \pi}} c_j e^{i\left(\omega-\omega_c\right) \tau} e^{-i \omega x_j / v}
\end{equation}
where we have taken $b_R (0)=0$ since the waveguide is initially in the vacuum state. On the other hand, the equation of motion of arbitrary operator $O$ is given by
\begin{widetext}
\begin{equation}
\frac{d}{d t} O(t)=\sum_{j=L, R} \int d \omega \sqrt{\frac{\kappa_c}{2 \pi}}\left\{b_R^{\dagger}(t) e^{i\left(\omega-\omega_c\right) t} e^{-i \omega x_j / v}\left[O(t), c_j(t)\right]-\left[O(t), c_j^{\dagger}(t)\right] b_R(t) e^{-i\left(\omega-\omega_c\right) t} e^{i \omega x_j / v}\right\}
\end{equation}
Substituting $b_R (t)$ into the above equation, we have
\begin{equation}\label{a8}
\frac{d}{d t} O(t)=\frac{\kappa_c}{2 \pi} \sum_{j, l=L, R} \int_0^t d \tau \int d \omega\left\{e^{i\left(\omega-\omega_c\right)(t-\tau)} e^{-i \omega x_j / v} c_l^{\dagger}(\tau)\left[O(t), c_j(t)\right]-\left[O(t), c_j^{\dagger}(t)\right] c_l(\tau) e^{-i\left(\omega-\omega_c\right)(t-\tau)} e^{i \omega x_{j l} / v}\right\}
\end{equation}
where $x_{jl}=x_j-x_l$, and we apply the Markov approximation by assuming the time delay $x_{jl}/v$ between the CCW mode and the mirrored CW mode can be neglected. Therefore,
\begin{equation}\label{a9}
\begin{gathered}
\frac{\kappa_c}{2 \pi} \sum_{l=L, R} \int_0^t d \tau \int d \omega e^{i\left(\omega-\omega_c\right)(t-\tau)} e^{-i \omega x_{j l} / v} c_l^{\dagger}(\tau)=\kappa_c \sum_{l=L, R} \int_0^t d \tau \delta\left(t-\frac{x_{j l}}{v}-\tau\right) e^{-i k x_{j l}} c_l^{\dagger}(\tau) \\
\approx \frac{\kappa_c}{2} c_j^{\dagger}(t)+\kappa_c \sum_{l=L, R} \Theta\left(t-\frac{x_{j l}}{v}\right) e^{-i k x_{j l}} c_l^{\dagger}(t)
\end{gathered}
\end{equation}
where $x_{jl}>0$ and $\Theta(t)$ is the step function. With Eq. (\ref{a9}) and taking the averages of Eq. (\ref{a8}), we have
\begin{equation}
\begin{aligned}
\frac{d}{d t}\langle O(t)\rangle=\frac{\kappa_c}{2} & \sum_{j=L, R}\left\{\left\langle c_j^{\dagger}(t)\left[O(t), c_j(t)\right]\right\rangle-\left\langle\left[O(t), c_j^{\dagger}(t)\right] c_j(t)\right\rangle\right\} \\
& +\kappa_c \sum_{j, l=L, R, j \neq l}\left\{e^{-i k x_{j l}}\left\langle c_l^{\dagger}(t)\left[O(t), c_j(t)\right]\right\rangle-e^{i k x_{j l}}\left\langle\left[O(t), c_j^{\dagger}(t)\right] c_l(t)\right\rangle\right\}
\end{aligned}
\end{equation}
Since $\langle O(t)\rangle=\operatorname{Tr}[O(t) \rho(0)]=\operatorname{Tr}[O \rho(t)]$, we can simplify the averages of operators in the above equation by using the cyclic property of trace
\begin{equation}
\left\langle c_\lambda^{\dagger}(t)\left[O(t), c_\lambda(t)\right]\right\rangle=\operatorname{Tr}\left[c_\lambda^{\dagger} O c_\lambda \rho(t)-c_\lambda^{\dagger} c_\lambda O \rho(t)\right]=\operatorname{Tr}\left[O c_\lambda \rho(t) c_\lambda^{\dagger}-O \rho(t) c_\lambda^{\dagger} c_\lambda\right]=\operatorname{Tr}\left\{O\left[c_\lambda, \rho(t) c_\lambda^{\dagger}\right]\right\}
\end{equation}
\begin{equation}
\left\langle\left[O(t), c_j^{\dagger}(t)\right] c_j(t)\right\rangle=\operatorname{Tr}\left[O c_j^{\dagger} c_j \rho(t)-c_j^{\dagger} O c_j \rho(t)\right]=\operatorname{Tr}\left[O c_j^{\dagger} c_j \rho(t)-O c_j \rho(t) c_j^{\dagger}\right]=\operatorname{Tr}\left\{O\left[c_j^{\dagger}, c_j \rho(t)\right]\right\}
\end{equation}
Therefore, we can obtain a QME in the following form 
\begin{equation}
\begin{gathered}
\frac{d}{d t} \rho(t)=-i[H, \rho(t)]+\frac{\kappa_c}{2} \sum_{j=c c w, c w}\left\{\left[c_j, \rho(t) c_j^{\dagger}\right]-\left[c_j^{\dagger}, c_j \rho(t)\right]\right\} \\
+\kappa_c \sum_{j, l=c c w, c w, j \neq l}\left\{e^{-i k x_{j l}}\left[c_j, \rho(t) c_l^{\dagger}\right]-e^{i k x_{j l}[}\left[c_j^{\dagger}, c_l \rho(t)\right]\right\}
\end{gathered}
\end{equation}
The second term on the right-hand side can be expanded and rewritten using the Liouvillian superoperator. By replacing $c_L$ and $c_R$ with $c_{ccw}$ and $c_{cw}$, respectively, and noting $kx_{jl}=\phi$ in the third term on the right-hand side, we arrive at the extended cascaded QME in Eqs. (\ref{eq1}) and (\ref{eq2}), except for the additional decay for individuals.

\section{Derivation of the complete spectral density of hybrid CEP cavity}\label{ab}

In the main text, we provide the component of modified plasmon $J_a (\omega)$ of spectral density, here we derive the analytical expressions of other components. For the modified cavity response $J_c(\omega)$, the dynamics of corresponding correlation functions is as follows: 
\begin{equation}
\frac{d}{d \tau}\left[\begin{array}{c}
\left\langle c_1^{\dagger}(0) a(\tau)\right\rangle \\
\left\langle c_1^{\dagger}(0) c_1(\tau)\right\rangle \\
\left\langle c_1^{\dagger}(0) c_2(\tau)\right\rangle
\end{array}\right]=-i\left[\begin{array}{ccc}
\omega_a-i \frac{\kappa_a}{2} & \sqrt{2} g_1 & 0 \\
\sqrt{2} g_1 & \omega_c-i \frac{\kappa_c\left(1+e^{i \phi}\right)}{2} & i \frac{\kappa_c}{2} e^{i \phi} \\
0 & -i \frac{\kappa_c}{2} e^{i \phi} & \omega_c-i \frac{\kappa_c\left(1-e^{i \phi}\right)}{2}
\end{array}\right]\left[\begin{array}{l}
\left\langle c_1^{\dagger}(0) a(\tau)\right\rangle \\
\left\langle c_1^{\dagger}(0) c_1(\tau)\right\rangle \\
\left\langle c_1^{\dagger}(0) c_2(\tau)\right\rangle
\end{array}\right]
\end{equation}
where the characteristic matrix takes the same form as $\mathbf{M_s}$ of Eq. (\ref{eq9}). With the initial conditions $\left\langle c_1^{\dagger}(0) a(0)\right\rangle=0$, $\left\langle c_1^{\dagger}(0) c_1(0)\right\rangle=1$ and $\left\langle c_1^{\dagger}(0) c_2(0)\right\rangle=0$, we obtain the following equation by taking the Laplace transform 
\begin{equation}
s \left[\begin{array}{c}
\left\langle c_1^{\dagger} a(s)\right\rangle \\
\left\langle c_1^{\dagger} c_1(s)\right\rangle \\
\left\langle c_1^{\dagger} c_2(s)\right\rangle
\end{array}\right]=-i\left[\begin{array}{ccc}
\omega_a-i \frac{\kappa_a}{2} & \sqrt{2} g_1 & 0 \\
\sqrt{2} g_1 & \omega_c-i \frac{\kappa_c\left(1+e^{i \phi}\right)}{2} & i \frac{\kappa_c}{2} e^{i \phi} \\
0 & -i \frac{\kappa_c}{2} e^{i \phi} & \omega_c-i \frac{\kappa_c\left(1-e^{i \phi}\right)}{2}
\end{array}\right]\left[\begin{array}{l}
\left\langle c_1^{\dagger} a(s)\right\rangle \\
\left\langle c_1^{\dagger} c_1(s)\right\rangle \\
\left\langle c_1^{\dagger} c_2(s)\right\rangle
\end{array}\right]+\left[\begin{array}{l}
0 \\
1 \\
0
\end{array}\right]
\end{equation}
The solution of $\left\langle c_1^{\dagger} c_1(\tau)\right\rangle$ is given by 
\begin{equation}\label{eqb3}
\left\langle c_1^{\dagger} c_1(\omega)\right\rangle=-\frac{i \chi_{c_1}(\omega)}{-1+2 g_1^2 \chi_a(\omega) \chi_{c_1}(\omega)+\left(\frac{\kappa_c}{2}\right)^2 e^{i 2 \phi} \chi_{c_1}(\omega) \chi_{c_2}(\omega)}
\end{equation}
where we have transformed from the $s$ domain to the frequency domain by replacing $s=-i \omega$, and introduced the polarizabilities for standing wave modes $c_1$ and $c_2$ 
\begin{equation}
\chi_{c_1}(\omega)=\frac{1}{\omega-\omega_c+i \frac{\kappa_c\left(1+e^{i \phi}\right)}{2}}
\end{equation}
\begin{equation}
\chi_{c_2}(\omega)=\frac{1}{\omega-\omega_c+i \frac{\kappa_c\left(1-e^{i \phi}\right)}{2}}
\end{equation}
We can rearrange terms in the denominator of Eq. (\ref{eqb3}) and obtain
\begin{equation}
\left\langle c_1^{\dagger} c_1(\omega)\right\rangle=\frac{i \chi_c(\omega)\left[1-i \frac{\kappa_c}{2} e^{i \phi} \chi_c(\omega)\right]}{1-g_1^2 \chi_a(\omega) \chi_{E P}(\omega)}
\end{equation}
Therefore, the modified cavity response is given by 
\begin{equation}\label{eqb7}
J_c(\omega)=-g_c^2 \operatorname{Im}\left[\chi_c(\omega) \frac{1-i \frac{\kappa_c}{2} e^{i \phi} \chi_c(\omega)}{1-g_1^2 \chi_a(\omega) \chi_{E P}(\omega)}\right]
\end{equation}
On the other hand, we can also obtain the solution of $\langle c_1^{\dagger} a(\omega)\rangle = \left[ \langle a^{\dagger} c_1(\omega)\rangle \right]^{\dagger}$ from Eq. (\ref{eq9}), which is
\begin{equation}
\left\langle c_1^{\dagger} a(\omega)\right\rangle=\frac{i \sqrt{2} g_1 \chi_a(\omega) \chi_{c_1}(\omega)}{1-2 g_1^2 \chi_a(\omega) \chi_{c_1}(\omega)-\left(\frac{\kappa_c}{2}\right)^2 e^{i 2 \phi} \chi_{c_1}(\omega) \chi_{c_2}(\omega)}
\end{equation}
After simplification, we obtain
\begin{equation}
\left\langle c_1^{\dagger} a(\omega)\right\rangle=\frac{i \sqrt{2} g_1 \chi_a(\omega) \chi_c(\omega)\left[1-i \frac{\kappa_c}{2} e^{i \phi} \chi_c(\omega)\right]}{1-g_1^2 \chi_a(\omega) \chi_{E P}(\omega)}
\end{equation}
which yields the crossing interference term of spectral density
\begin{equation}\label{eqb10}
J_{a c}(\omega)=-2 g_a g_c \operatorname{Im}\left[g_1 \chi_a(\omega) \chi_c(\omega) \frac{1-i \frac{\kappa_c}{2} e^{i \phi} \chi_c(\omega)}{1-g_1^2 \chi_a(\omega) \chi_{E P}(\omega)}\right]
\end{equation}
We can see that $J_c (\omega)$ and $J_{ac} (\omega)$ exhibit more complex dependence on $\phi$ compared to $J_a (\omega)$. In the case of far red-detuned plasmon-photon coupling, it is possible to derive a simpler analytical expression of the spectral density by adiabatically eliminating the equation of plasmonic mode. The technical details of this approximation can be referred to Ref. \cite{RN18}. 

The complete spectral density is given by 
\begin{equation}\label{eqb11}
\begin{aligned}
J(\omega) & =J_a (\omega)+J_c (\omega)+J_{ac} (\omega) \\ 
& =-\operatorname{Im}\left\{\frac{g_a^2 \chi_a(\omega)+g_c \chi_c(\omega)\left[1-i \frac{\kappa_c}{2} e^{i \phi} \chi_c(\omega)\right]\left[g_c+2 g_a g_1 \chi_a(\omega)\right]}{1-g_1^2 \chi_a(\omega) \chi_{E P}(\omega)}\right\}
\end{aligned}
\end{equation}

\section{Derivation of the emission spectrum of QE}\label{ac}

The emission spectrum of QE studied here is also called the spontaneous emission spectrum or the polarization spectrum \cite{RN69}, which reflects the local dynamics of a QE. We express the emission spectrum of QE in term of the polarizability of hybrid CEP cavity for the sake of physical transparency. The correlation $\left\langle\sigma_{+}(t) \sigma_{-}(t+\tau)\right\rangle$ can be calculated from Eqs. (\ref{eq1}) and (\ref{eq2}) using the quantum regression theorem \cite{RN50}, which yields the following equation of motion 
\begin{equation}\label{eqc1}
i \frac{d}{d \tau}\left[\begin{array}{c}
\left\langle\sigma_{+}(0) \sigma_{-}(\tau)\right\rangle \\
\left\langle\sigma_{+}(0) a(\tau)\right\rangle \\
\left\langle\sigma_{+}(0) c_1(\tau)\right\rangle \\
\left\langle\sigma_{+}(0) c_2(\tau)\right\rangle
\end{array}\right]=\mathbf{M}_p\left[\begin{array}{c}
\left\langle\sigma_{+}(0) \sigma_{-}(\tau)\right\rangle \\
\left\langle\sigma_{+}(0) a(\tau)\right\rangle \\
\left\langle\sigma_{+}(0) c_1(\tau)\right\rangle \\
\left\langle\sigma_{+}(0) c_2(\tau)\right\rangle
\end{array}\right]
\end{equation}
where the matrix $\mathbf{M}_p$ is given in Eq. (\ref{eq6}). The initial condition is 
\begin{equation}
\left[\begin{array}{c}
\left\langle\sigma_{+}(0) \sigma_{-}(0)\right\rangle \\
\left\langle\sigma_{+}(0) a(0)\right\rangle \\
\left\langle\sigma_{+}(0) c_1(0)\right\rangle \\
\left\langle\sigma_{+}(0) c_2(0)\right\rangle
\end{array}\right]=\left[\begin{array}{l}
1 \\
0 \\
0 \\
0
\end{array}\right]
\end{equation}
Laplace transforming Eq. (\ref{eqc1}) into the $s$ domain, we have
\begin{equation}\label{eqc3}
\left[s+i\left(\omega_0-i \frac{\gamma}{2}\right)\right]\left\langle\sigma_{+} \sigma_{-}(s)\right\rangle=1-i g_a\left\langle\sigma_{+} a(s)\right\rangle-i \sqrt{2} g_c\left\langle\sigma_{+} c_1(s)\right\rangle
\end{equation}
The remaining correlation functions can be solved by the following equation
\begin{equation}
s\left[\begin{array}{c}
\left\langle\sigma_{+} a(s)\right\rangle \\
\left\langle\sigma_{+} c_1(s)\right\rangle \\
\left\langle\sigma_{+} c_2(s)\right\rangle
\end{array}\right]=-i\left[\begin{array}{ccc}
\omega_a-i \frac{\kappa_a}{2} & \sqrt{2} g_1 & 0 \\
\sqrt{2} g_1 & \omega_c-i \frac{\kappa_c\left(1+e^{i \phi}\right)}{2} & i \frac{\kappa_c}{2} e^{i \phi} \\
0 & -i \frac{\kappa_c}{2} e^{i \phi} & \omega_c-i \frac{\kappa_c\left(1-e^{i \phi}\right)}{2}
\end{array}\right]\left[\begin{array}{c}
\left\langle\sigma_{+} a(s)\right\rangle \\
\left\langle\sigma_{+} c_1(s)\right\rangle \\
\left\langle\sigma_{+} c_2(s)\right\rangle
\end{array}\right]-i\left[\begin{array}{c}
g_a \\
\sqrt{2} g_c \\
0
\end{array}\right]\left\langle\sigma_{+} \sigma_{-}(s)\right\rangle
\end{equation}
We can see from Eq. (\ref{eqc3}) that only the correlation functions $\left\langle\sigma_{+} a(s)\right\rangle$ and $\left\langle\sigma_{+} c_1(s)\right\rangle$ are needed, which are given by 
\begin{equation}
\left\langle\sigma_{+} a(s)\right\rangle=-i\left[g_a\left\langle a^{\dagger} a(s)\right\rangle+\sqrt{2} g_c\left\langle a^{\dagger} c_1(s)\right\rangle\right]\left\langle\sigma_{+} \sigma_{-}(s)\right\rangle
\end{equation}
\begin{equation}
\left\langle\sigma_{+} c_1(s)\right\rangle=-i\left[\sqrt{2} g_c\left\langle c_1^{\dagger} c_1(s)\right\rangle+g_a\left\langle c_1^{\dagger} a(s)\right\rangle\right]\left\langle\sigma_{+} \sigma_{-}(s)\right\rangle
\end{equation}
Substituting the above equations into Eq. (\ref{eqc3}), we obtain
\begin{equation}
\left[s+i\left(\omega_0-i \frac{\gamma}{2}\right)+\left\langle\Lambda^{\dagger} \Lambda(s)\right\rangle\right]\left\langle\sigma_{+} \sigma_{-}(s)\right\rangle=1
\end{equation}
where $\Lambda$ is given in Eq. (\ref{eq8}). Transforming into the frequency domain by replacing $s=-i\omega$, we have 
\begin{equation}\label{eqc8}
\left\langle\sigma_{+} \sigma_{-}(\omega)\right\rangle=\frac{i}{\omega-\omega_0+i \frac{\gamma}{2}-\chi_{s y s}(\omega)}
\end{equation}
where $\chi_{s y s}(\omega)$ is defined as the cavity polarizability in Eq. (\ref{eq7}). Eq. (\ref{eqc8}) can be rewritten as 
\begin{equation}
\left\langle\sigma_{+} \sigma_{-}(\omega)\right\rangle=\frac{i}{\omega-\omega_0-\Delta(\omega)+i \frac{\gamma+\Gamma(\omega)}{2}}
\end{equation}
with the photon induced Lamb shift $\Delta(\omega)=\operatorname{Re}\left[\chi_{\text {sys }}(\omega)\right]$ and the local coupling strength $\Gamma(\omega)=-2 \operatorname{Im}\left[\chi_{s y s}(\omega)\right]$. Therefore, the emission spectrum of QE is expressed as 
\begin{equation}
S(\omega)=\frac{1}{\pi} \frac{\gamma+\Gamma(\omega)}{\left[\omega-\omega_0-\Delta(\omega)\right]^2+\left[\frac{\gamma+\Gamma(\omega)}{2}\right]^2}
\end{equation}

\section{Eigenmode decomposition of the scattering spectrum of hybrid CEP cavity}\label{ad}

In this section, we present a formalism that the cavity scattering is described by the radiation of individual superradiant and subradiant eigenresonances and their interferences, thus allows to identify the contributions of eigenmodes to the scattering spectrum of hybrid CEP cavity and explain the exotic radiation enhancement of plasmonic-photonic cavity operating at CEP with $\phi=0$. We start by implementing a driving Hamiltonian for plasmon driven case in Eq. (\ref{eq1}), which reads 
\begin{equation}
H_p=q_{i n}\left(e^{-i \omega_L t} a^{\dagger}+a e^{i \omega_L t}\right)
\end{equation}
where $\omega_L$ is the frequency of laser field and $q_{in}$ is the driving strength. Applying the unitary transformation $U=\exp \left[-i \omega_L\left(c_{c c w}^{\dagger} c_{c c w}+c_{c w}^{\dagger} c_{c w}+a^{\dagger} a+\sigma_{+} \sigma_{-}\right) t\right]$, we can obtain the following equations of motion in the basis of standing wave modes 
\begin{equation}\label{eqd2}
\dot{a}=-i\left(\Delta_L-i \frac{\kappa_a}{2}\right) a-i \sqrt{2} g_1 c_1-i q_{in}
\end{equation}
\begin{equation}
\dot{c}_1=-i\left[\Delta_L-i \frac{\kappa_c\left(1+e^{i \phi}\right)}{2}\right] c_1-i \sqrt{2} g_1 a-\frac{\kappa_c}{2} e^{i \phi} c_2
\end{equation}
\begin{equation}\label{eqd4}
\dot{c}_2=-i\left[\Delta_L-i \frac{\kappa_c\left(1-e^{i \phi}\right)}{2}\right] c_2+\frac{\kappa_c}{2} e^{i \phi} c_1
\end{equation}
where we assume the resonant plasmon-photon coupling $\omega_c=\omega_a$, thus $\Delta_L=\omega-\omega_c=\omega-\omega_a$. By defining $\vec{c}=\left[a,c_1,c_2 \right]^T$, we can rewrite the above equations as 
\begin{equation}\label{eqd5}
i \frac{d \vec{c}}{d t}=V^{-1} B V \vec{c}+s_p
\end{equation}
where $B$ is the diagonal matrix formed from the eigenvalues of the characteristic matrix of Eqs. (\ref{eqd2})-(\ref{eqd4}) 
\begin{equation}
B=\left[\begin{array}{ccc}
\omega_a+i \gamma_a & 0 & 0 \\
0 & \omega_b+i \gamma_b & 0 \\
0 & 0 & \omega_c+i \gamma_c
\end{array}\right]
\end{equation}
and $V$ specifies the matrix where the rows are constituted by the corresponding left eigenvectors
\begin{equation}
V=\left[\begin{array}{lll}
v_{a, 1} & v_{a, 2} & v_{a, 3} \\
v_{b, 1} & v_{b, 2} & v_{b, 3} \\
v_{c, 1} & v_{c, 2} & v_{c, 3}
\end{array}\right]
\end{equation}
and $s_p=\left[q_{in},0,0\right]^T$ is the vector for input fields. Eq. (\ref{eqd5}) can be formally solved by using the Fourier transform 
\begin{equation}
\vec{c}\left(\Delta_L\right)=V^{-1}\left(\Delta_L I-B\right)^{-1} V s_p
\end{equation}
Therefore, we have
\begin{equation}
s\left(\Delta_L\right)=K \vec{c}\left(\Delta_L\right)=K V^{-1}\left(\Delta_L I-B\right)^{-1} V s_p
\end{equation}
where $K$ defines a matrix describing the coupling between different radiation channels
\begin{equation}
\Gamma=K^{\dagger} K=\left[\begin{array}{ccc}
\kappa_a & 0 & 0 \\
0 & \kappa_c & 0 \\
0 & 0 & \kappa_c
\end{array}\right]
\end{equation}
The scattering spectrum is given by
\begin{equation}
\sigma\left(\Delta_L\right)=s^{\dagger}\left(\Delta_L\right) s\left(\Delta_L\right)=\left[K V^{-1}\left(\Delta_L I-B\right)^{-1} V s_p\right]^{\dagger}\left[K V^{-1}\left(\Delta_L I-B\right)^{-1} V s_p\right]
\end{equation}
We introduce a matrix $G$ of imaginary eigenenergies 
\begin{equation}
G=\left[\begin{array}{ccc}
\gamma_a & 0 & 0 \\
0 & \gamma_b & 0 \\
0 & 0 & \gamma_c
\end{array}\right]
\end{equation}
Then the scattering spectrum can be written as
\begin{equation}\label{eqd13}
\begin{aligned}
\sigma\left(\Delta_L\right) & =\left[K V^{-1} G^{-1} G\left(\Delta_L I-B\right)^{-1} V s_p\right]^{\dagger}\left[K V^{-1} G^{-1} G\left(\Delta_L I-B\right)^{-1} V s_p\right] \\
& =\left[G\left(\Delta_L I-B\right)^{-1} V s_p V^{-1} G^{-1}\right]^{\dagger} \Gamma V^{-1} G^{-1} G\left(\Delta_L I-B\right)^{-1} V s_p
\end{aligned}
\end{equation}
where $G(\Delta_L I-B)^{-1}$ yields the Lorentzian lineshape
\begin{equation}
G\left(\Delta_L I-B\right)^{-1}=\left[\begin{array}{ccc}
\frac{\gamma_a}{\Delta_L-\omega_a-i \gamma_a} & 0 & 0 \\
0 & \frac{\gamma_b}{\Delta_L-\omega_b-i \gamma_b} & 0 \\
0 & 0 & \frac{\gamma_c}{\Delta_L-\omega_c-i \gamma_c}
\end{array}\right]
\end{equation}
$Vs_p$ gives the complex radiation patterns
\begin{equation}
V s_p=\left[\begin{array}{l}
v_{a, 1} \\
v_{b, 1} \\
v_{c, 1}
\end{array}\right]=\left[\begin{array}{l}
C_a \\
C_b \\
C_c
\end{array}\right]
\end{equation}
The remaining part of Eq. (\ref{eqd13}) can be simplified as
\begin{equation}\label{eqd16}
\begin{gathered}
\left[V^{-1} G^{-1}\right]^{\dagger} \Gamma V^{-1} G^{-1}=\frac{1}{\operatorname{Det}[V]^2}\left\{\left[\begin{array}{lll}
v_{b, 2} v_{c, 3}-v_{b, 3} v_{c, 2} & v_{a, 3} v_{c, 2}-v_{a, 2} v_{c, 3} & v_{a, 2} v_{b, 3}-v_{a, 3} v_{b, 2} \\
v_{b, 3} v_{c, 1}-v_{b, 1} v_{c, 3} & v_{a, 1} v_{c, 3}-v_{a, 3} v_{c, 1} & v_{a, 3} v_{b, 1}-v_{a, 1} v_{b, 3} \\
v_{b, 1} v_{c, 2}-v_{b, 2} v_{c, 1} & v_{a, 2} v_{c, 1}-v_{a, 1} v_{c, 2} & v_{a, 1} v_{b, 2}-v_{a, 2} v_{b, 1}
\end{array}\right]\left[\begin{array}{ccc}
\frac{1}{\gamma_a} & 0 & 0 \\
0 & \frac{1}{\gamma_b} & 0 \\
0 & 0 & \frac{1}{\gamma_c}
\end{array}\right]\right\}^{\dagger}\\
\times\left[\begin{array}{ccc}
\kappa_a & 0 & 0 \\
0 & \kappa_c & 0 \\
0 & 0 & \kappa_c
\end{array}\right]\left[\begin{array}{lll}
v_{b, 2} v_{c, 3}-v_{b, 3} v_{c, 2} & v_{a, 3} v_{c, 2}-v_{a, 2} v_{c, 3} & v_{a, 2} v_{b, 3}-v_{a, 3} v_{b, 2} \\
v_{b, 3} v_{c, 1}-v_{b, 1} v_{c, 3} & v_{a, 1} v_{c, 3}-v_{a, 3} v_{c, 1} & v_{a, 3} v_{b, 1}-v_{a, 1} v_{b, 3} \\
v_{b, 1} v_{c, 2}-v_{b, 2} v_{c, 1} & v_{a, 2} v_{c, 1}-v_{a, 1} v_{c, 2} & v_{a, 1} v_{b, 2}-v_{a, 2} v_{b, 1}
\end{array}\right]\left[\begin{array}{ccc}
\frac{1}{\gamma_a} & 0 & 0 \\
0 & \frac{1}{\gamma_b} & 0 \\
0 & 0 & \frac{1}{\gamma_c}
\end{array}\right]\\
=p\left[\begin{array}{cccc}
\frac{\left|V_{1,1}^{-1}\right|^2 \kappa_a+\left(\left|V_{2,1}^{-1}\right|^2+\left|V_{3,1}^{-1}\right|^2\right) \kappa_c}{\gamma_a^2} & \frac{V_{1,1}^{-1 *} V_{1,2}^{-1} \kappa_a+\left(V_{2,1}^{-1 *} V_{2,2}^{-1}+V_{3,1}^{-1 *} V_{3,2}^{-1}\right) \kappa_c}{\gamma_a \gamma_b} & \frac{V_{1,1}^{-1 *} V_{1,3}^{-1} \kappa_a+\left(V_{2,1}^{-1 *} V_{2,3}^{-1}+V_{3,1}^{-1 *} V_{3,3}^{-1}\right) \kappa_c}{\gamma_a \gamma_c} \\
\frac{V_{1,2}^{-1 *} V_{1,1}^{-1} \kappa_a+\left(V_{2,2}^{-1 *} V_{2,1}^{-1}+V_{3,2}^{-1 *} V_{3,1}^{-1}\right) \kappa_c}{\gamma_a \gamma_b} & \frac{\left|V_{1,2}^{-1}\right|^2 \kappa_a+\left(\left|V_{2,2}^{-1}\right|^2+\left|V_{3,2}^{-1}\right|^2\right) \kappa_c}{\gamma_b^2} & \frac{V_{1,2}^{-1 *} V_{1,3}^{-1} \kappa_a+\left(V_{2,2}^{-1 *} V_{2,3}^{-1}+V_{3,2}^{-1 *} V_{3,3}^{-1}\right) \kappa_c}{\gamma_b \gamma_c} \\
\frac{V_{1,3}^{-1 *} V_{1,1}^{-1} \kappa_a+\left(V_{2,3}^{-1 *} V_{2,1}^{-1}+V_{3,3}^{-1 *} V_{3,1}^{-1}\right) \kappa_c}{\gamma_a \gamma_c} & \frac{V_{1,3}^{-1 *} V_{1,2}^{-1} \kappa_a+\left(V_{2,3}^{-1 *} V_{2,2}^{-1}+V_{3,3}^{-1 *} V_{3,2}^{-1}\right) \kappa_c}{\gamma_b \gamma_c} & \frac{\left|V_{1,3}^{-1}\right|^2 \kappa_a+\left(\left|V_{2,3}^{-1}\right|^2+\left|V_{3,3}^{-1}\right|^2\right) \kappa_c}{\gamma_c^2}
\end{array}\right]
\end{gathered}
\end{equation}
where $p=\operatorname{Det}[V]^{-2}$ and $V_{i,j}^{-1}$ indexes the elements of matrix $V^{-1}$ 
\begin{equation}
V^{-1}=\frac{1}{\operatorname{Det}[V]}\left[\begin{array}{lll}v_{b, 2} v_{c, 3}-v_{b, 3} v_{c, 2} & v_{a, 3} v_{c, 2}-v_{a, 2} v_{c, 3} & v_{a, 2} v_{b, 3}-v_{a, 3} v_{b, 2} \\ v_{b, 3} v_{c, 1}-v_{b, 1} v_{c, 3} & v_{a, 1} v_{c, 3}-v_{a, 3} v_{c, 1} & v_{a, 3} v_{b, 1}-v_{a, 1} v_{b, 3} \\ v_{b, 1} v_{c, 2}-v_{b, 2} v_{c, 1} & v_{a, 2} v_{c, 1}-v_{a, 1} v_{c, 2} & v_{a, 1} v_{b, 2}-v_{a, 2} v_{b, 1}\end{array}\right]
\end{equation}
Therefore, Eq. (\ref{eqd16}) can be expressed in a compact form
\begin{equation}
\left[V^{-1} G^{-1}\right]^{\dagger} \Gamma V^{-1} G^{-1}=\frac{p}{\gamma_a \gamma_b \gamma_c}\left[\begin{array}{ccc}
h_{a a} & h_{a b} & h_{a c} \\
h_{a b}^* & h_{b b} & h_{b c} \\
h_{a c}^* & h_{b c}^* & h_{c c}
\end{array}\right]
\end{equation}
with 
\begin{equation}
\begin{gathered}
h_{a b}=\gamma_c\left[V_{1,1}^{-1 *} V_{1,2}^{-1} \kappa_a+\left(V_{2,1}^{-1 *} V_{2,2}^{-1}+V_{3,1}^{-1 *} V_{3,2}^{-1}\right) \kappa_c\right] \\
h_{a c}=\gamma_b\left[V_{1,1}^{-1 *} V_{1,3}^{-1} \kappa_a+\left(V_{2,1}^{-1 *} V_{2,3}^{-1}+V_{3,1}^{-1 *} V_{3,3}^{-1}\right) \kappa_c\right] \\
h_{b c}=\gamma_a\left[V_{1,2}^{-1 *} V_{1,3}^{-1} \kappa_a+\left(V_{2,2}^{-1 *} V_{2,3}^{-1}+V_{3,2}^{-1 *} V_{3,3}^{-1}\right) \kappa_c\right] \\
h_{a a}=\frac{\gamma_b \gamma_c}{\gamma_a}\left[\left|V_{1,1}^{-1}\right|^2 \kappa_a+\left(\left|V_{2,1}^{-1}\right|^2+\left|V_{3,1}^{-1}\right|^2\right) \kappa_c\right] \\
h_{b b}=\frac{\gamma_a \gamma_c}{\gamma_b}\left[\left|V_{1,2}^{-1}\right|^2 \kappa_a+\left(\left|V_{2,2}^{-1}\right|^2+\left|V_{3,2}^{-1}\right|^2\right) \kappa_c\right] \\
h_{c c}=\frac{\gamma_a \gamma_b}{\gamma_c}\left[\left|V_{1,3}^{-1}\right|^2 \kappa_a+\left(\left|V_{2,3}^{-1}\right|^2+\left|V_{3,3}^{-1}\right|^2\right) \kappa_c\right]
\end{gathered}
\end{equation}
Accordingly, the scattering spectrum can be further simplified as
\begin{equation}
\begin{aligned}
\sigma\left(\Delta_L\right) & =\frac{p}{\gamma_a \gamma_b \gamma_c}\left[\begin{array}{lll}
C_a^* & C_b^* & C_c^*
\end{array}\right]\left[\begin{array}{ccc}
\frac{\gamma_a}{\Delta_L-\omega_a+i \gamma_a} & 0 & 0 \\
0 & \frac{\gamma_b}{\Delta_L-\omega_b+i \gamma_b} & 0 \\
0 & 0 & \frac{\gamma_c}{\Delta_L-\omega_c+i \gamma_c}
\end{array}\right]\left[\begin{array}{ccc}
h_{a a} & h_{a b} & h_{a c} \\
h_{a b}^* & h_{b b} & h_{b c} \\
h_{a c}^* & h_{b c}^* & h_{c c}
\end{array}\right] \\
& \qquad\qquad\qquad\qquad\qquad \times\left[\begin{array}{ccc}
\frac{\gamma_a}{\Delta_L-\omega_a-i \gamma_a} & 0 & 0 \\
0 & \frac{\gamma_b}{\Delta_L-\omega_b-i \gamma_b} & 0 \\
0 & 0 & \frac{\gamma_c}{\Delta_L-\omega_c-i \gamma_c} 
\end{array}\right] \left[\begin{array}{c}
C_a \\
C_b \\
C_c
\end{array}\right]\\
& = \frac{p}{\gamma_a \gamma_b \gamma_c}\left[\begin{array}{c}
\frac{h_{a a} C_a^* \gamma_a^2}{\left(\Delta_L-\omega_a\right)^2+\gamma_a^2}+\frac{h_{a b}^* C_b^* \gamma_a \gamma_b}{\left(\Delta_L-\omega_a-i \gamma_a\right)\left(\Delta_L-\omega_b+i \gamma_b\right)}+\frac{h_{a c}^* C_c^* \gamma_a \gamma_c}{\left(\Delta_L-\omega_a-i \gamma_a\right)\left(\Delta_L-\omega_c+i \gamma_c\right)} \\
\frac{h_{a b} C_a^* \gamma_a \gamma_b}{\left(\Delta_L-\omega_a+i \gamma_a\right)\left(\Delta_L-\omega_b-i \gamma_b\right)}+\frac{h_{b b} C_b^* \gamma_b^2}{\left(\Delta_L-\omega_b\right)^2+\gamma_b^2}+\frac{h_{b c}^* C_c^* \gamma_b \gamma_c}{\left(\Delta_L-\omega_b-i \gamma_b\right)\left(\Delta_L-\omega_c+i \gamma_c\right)} \\
\frac{h_{a c} C_a^* \gamma_a \gamma_c}{\left(\Delta_L-\omega_a+i \gamma_a\right)\left(\Delta_L-\omega_c-i \gamma_c\right)}+\frac{h_{b c} C_b^* \gamma_b \gamma_c}{\left(\Delta_L-\omega_b+i \gamma_b\right)\left(\Delta_L-\omega_c-i \gamma_c\right)}+\frac{h_{c c} C_c^* \gamma_c^2}{\left(\Delta_L-\omega_c\right)^2+\gamma_c^2}
\end{array}\right]^T\left[\begin{array}{c}
C_a \\
C_b \\
C_c
\end{array}\right]
\end{aligned}
\end{equation}
Finally, we arrive at
\begin{equation}
\begin{gathered}
\sigma\left(\Delta_L\right)=p_{\gamma}\left\{\frac{h_{a a}\left|C_a\right|^2 \gamma_a^2}{\left(\Delta_L-\omega_a\right)^2+\gamma_a^2}+\frac{h_{b b}\left|C_b\right|^2 \gamma_b^2}{\left(\Delta_L-\omega_b\right)^2+\gamma_b^2}+\frac{h_{c c}\left|C_c\right|^2 \gamma_c^2}{\left(\Delta_L-\omega_c\right)^2+\gamma_c^2}+2 \operatorname{Re}\left[\frac{h_{a b} C_a^* C_b \gamma_a \gamma_b}{\left(\Delta_L-\omega_a+i \gamma_a\right)\left(\Delta_L-\omega_b-i \gamma_b\right)}\right]\right. \\
\left.+2 \operatorname{Re}\left[\frac{h_{a c} C_a^* C_c \gamma_a \gamma_c}{\left(\Delta_L-\omega_a+i \gamma_a\right)\left(\Delta_L-\omega_c-i \gamma_c\right)}\right]+2 \operatorname{Re}\left[\frac{h_{bc} C_b^* C_c \gamma_b \gamma_c}{\left(\Delta_L-\omega_b+i \gamma_b\right)\left(\Delta_L-\omega_c-i \gamma_c\right)}\right]\right\}
\end{gathered}
\end{equation}
where $p_{\gamma}=p/\gamma_a \gamma_b \gamma_c$. At $\Delta_L=\omega_a=\omega_b=\omega_c$, the scattering intensity is given by 
\begin{equation}
\sigma(0)=p_{\gamma}\left\{h_{a a}\left|C_a\right|^2+h_{b b}\left|C_b\right|^2+h_{c c}\left|C_c\right|^2+2 \operatorname{Re}\left[h_{a b} C_a^* C_b+h_{a c} C_a^* C_c+h_{b c} C_b^* C_c\right]\right\}
\end{equation}

\end{widetext}

\nocite{*}

\bibliography{LDOS}

\end{document}